\documentclass[lettersize,journal]{IEEEtran}
\usepackage{amsmath,amsfonts}
\usepackage{algorithmic}
\usepackage{algorithm}
\usepackage{array}
\usepackage[caption=false,font=normalsize,labelfont=sf,textfont=sf]{subfig}
\usepackage{textcomp}
\usepackage{stfloats}
\usepackage{url}
\usepackage{verbatim}
\usepackage{graphicx}
\usepackage{cite}
\usepackage{float} 
\usepackage{subcaption}
 \usepackage{multirow}
 \usepackage{caption}
\usepackage{threeparttable}
\usepackage{amsthm}
\usepackage{arydshln}
\usepackage{amssymb}
\usepackage{amsmath}
\usepackage{longtable}
\usepackage{array}
\usepackage{graphicx}
 \usepackage{multirow}
\usepackage{color}
\usepackage{enumerate}
\usepackage{booktabs}
\usepackage{longtable}
\usepackage[latin1]{inputenc}

\newtheorem{theorem}{\bf Theorem}
\newtheorem{lemma}{Lemma}

\newtheorem{example}{Example}

\newtheorem{remark}{Remark}

\newcommand{\tr}{{\mathrm{Tr}}}

\newcommand{\gf}{{\mathbb{F}}}
 \allowdisplaybreaks[3]
\begin{document}

\title{New constructions of asymptotically optimal periodic and aperiodic quasi-complementary sequence sets}

\author{ Peng Wang, Ziling Heng and Chengju Li \thanks{P. Wang and Z. Heng are with the School of Science, Chang'an University, Xi'an 710064, China,  and also with the National Mobile Communications Research Laboratory, Southeast University, Nanjing 211111, China (email: wp20201115@163.com, zilingheng@chd.edu.cn). C. Li is with the MoE Engineering Research Center of Software/Hardware Co-design Technology and Application, East China Normal University, Shanghai, 200062, China (email: cjli@sei.ecnu.edu.cn).
(\emph{Corresponding author: Ziling Heng})}
\thanks{Z. Heng's research was supported in part by the National Natural Science Foundation of China under Grant 12271059, in part by the Shaanxi Fundamental
 Science Research Project for Mathematics and Physics (Grant No. 23JSZ008), in part by the open research fund of National Mobile Communications 
 Research Laboratory of Southeast University under Grant 2024D10 and in part by the Research Funds for the Interdisciplinary Projects, CHU, under Grant 300104240922.
  C. Li's research was supported in part by the National Natural Science Foundation of China (Grant No. T2322007), in part by the Shanghai
Rising-Star Program (Grant No. 22QA1403200), and in part by the Shanghai Natural Science Foundation (Grant No. 22ZR1419600).
 }}

\maketitle

\begin{abstract}
Quasi-complementary sequence sets (QCSSs) play an important role in  multi-carrier code division multiple access (MC-CDMA) systems as they can support more users than perfect complementary sequence sets (PCSSs).  The objective of this paper is to present new constructions of asymptotically optimal periodic and aperiodic  QCSSs with large set sizes. Firstly, we construct a family of  asymptotically optimal periodic $(p^{2n}, p^n-1, p^n-1, p^n+1)$ QCSSs with small alphabet size $p$,  which has larger set size than the known family of  periodic $(p^n(p^n-1), p^n-1, p^n-1, p^n+1)$ QCSSs. Secondly, we construct five new families of asymptotically optimal aperiodic QCSSs with large set sizes and low aperiodic tolerances.
Each family of these aperiodic QCSSs has set size $\Theta(K^2)$ for some flock size $K$.  Compared with known asymptotically optimal aperiodic QCSSs in the literature, the proposed aperiodic QCSSs by us have better parameters or new lengths of their constituent sequences.
\end{abstract}


\section{Introduction}
A perfect complementary sequence set (PCSS) is a collection of two-dimensional matrices with zero non-trivial auto-correlation sums and zero cross-correlation sums for all time shifts.
The rows of each matrix in the PCSS are called  constituent sequences of the matrix.  When the PCSS has only one matrix and this matrix has only two rows, then the two constituent sequences of the matrix compose a Golay complementary pair \cite{G}.  As PCSSs have ideal correlation properties, they are widely applied in multi-carrier code division multiple access (MC-CDMA) systems  for decreasing the peak-to-average power ratio \cite{JA1}, radar waveform design\cite{AP1} and channel estimation\cite{PS1,SW1}. In an MC-CDMA communication system, each user is assigned a two-dimensional matrix and all of its constituent sequences are transmitted over separate sub-carrier channels simultaneously \cite{C1}.  Then the number of constituent sequences of a matrix in the PCSS is equal to the number of sub-carrier channels. 
The flock size of the PCSS is defined as the number of constituent sequences of each matrix. 
It is known that the set size of a PCSS is upper bounded by its flock size \cite{ZL}.  The PCSS achieving this bound is called a complete complementary code.

In order to support more users in MC-CDMA systems, several generalizations of PCSSs were proposed in the literature.
In \cite{T}, Tang, Fan, Li and Suehiro proposed the zero-correlation zone complementary sequence set (ZCZ-CSS) which has zero correlation sums within a zone around the in-phase position. The ZCZ-CSS is a generalization of the well-known Z-complementary sequence sets defined in \cite{F} from one-dimensional sequences to two-dimensional matrices. 
Later, Liu, Guan, Ng and Chen introduced the extension of ZCZ-CSS which is called the low-correlation zone complementary sequence set (LCZ-CSS) \cite{ZL2}.
The correlation sum of the LCZ-CSS is required to be  small  within a zone around the in-phase position.
If the correlation sum equals zero, then the LCZ-CSS becomes the ZCZ-CSS. LCZ-CSSs have larger set sizes than ZCSSs at the cost of correlation sums by the bounds given in \cite{ZL2}.
Nevertheless, the out-of-zone correlation sums of  LCZ-CSSs and LCZ-CSSs may be large, which makes them unsuitable for asynchronous MC-CDMA systems.
Then Liu et al. proposed the quasi-complementary sequence sets (QCSSs) which are appropriate for asynchronous MC-CDMA systems \cite{Liu, ZL}.
QCSSs are required to have small maximum non-trivial correlation sums for all non-zero time shifts. Clearly, ZCZ-CSSs and LCZ-CSSs are two special types of QCSSs. 
In recent years, the QCSS has attracted much attention due to its large set size and low interference performance\cite{ZL,ZL3}.
A QCSS involves four important parameters including the set size $M$, the flock size $K$, the length $N$ of each constituent sequence and the maximum correlation sum $\vartheta_{\max}$.
For fixed $K$ and $N$, it is desirable to construct an $(M,K,N,\vartheta_{\max})$-QCSS over an alphabet such that $M$ ($>K$) is as large as possible and  $\vartheta_{\max}$ is as small as possible. However, there exists a tradeoff among these parameters. In \cite{ZL}, Liu et al. gave the lower bound for $\vartheta_{\max}$ of a periodic $(M,K,N,\vartheta_{\max})$-QCSS.
In \cite{Welch}, Welch presented a correlation lower bound for an aperiodic $(M, K ,N, \vartheta_{\max})$-QCSS. 
In \cite{ZL1, ZL4}, Liu et al. established some tighter correlation lower bounds for an aperiodic $(M, K ,N, \vartheta_{\max})$-QCSS. Recently, Sarkar et al. derived a new  aperiodic correlation lower bound for a type of QCSSs, which is tighter than the existing bounds for QCSSs \cite{PS2}.

The optimality of a QCSS with respect to a lower bound is usually  evaluated by the optimality factor $\rho$, which is defined as the ratio of its maximum correlation magnitude and the lower bound \cite{ZL}. An $(M, K ,N, \vartheta_{\max})$-QCSS is said to be optimal if $\rho=1$, near-optimal if $1<\rho\leq 2$ and  asymptotically optimal if $\lim_{N\rightarrow +\infty}\rho=1$. Up to now, no optimal $(M, K ,N, \vartheta_{\max})$-QCSS with $M>K$ was reported and only a few families of asymptitically optimal or near optimal QCSSs were constructed in the literature. For periodic QCSSs, Liu et al. presented the first family of asymptotically optimal and the first family of near optimal QCSSs by using Singer difference sets \cite{ZL}. 
Following this work, some families of asymptotically optimal periodic QCSSs were constructed from almost difference sets, cyclic difference sets, cyclotomic classes and additive characters over finite fields \cite{LY1,LY2,LY3,LY4, LG, XLC}. Most recently, Heng et al. used characters and polynomials over finite fields to construct several families of asymptotically optimal periodic QCSSs with set size $\Theta(K^2)$ and near optimal periodic QCSSs with set size $\Theta(K^3)$, which have better parameters than known ones \cite{ZH1}. 
Besides, there also exist some known constructions of asymptotically optimal aperiodic QCSSs in the literature. 
In \cite{LY5}, Li et al. constructed three families of asymptotically optimal aperiodic QCSSs. In \cite{ZZ1, ZZ2,ZZ3}, based on several algebraic tools including permutation functions,  extended Boolean functions and Florentine rectangles, various families of asymptotically optimal and near-optimal aperiodic QCSSs were constructed.  
Recently, Sarkar et al. presented a systematic construction of asymptotically optimal aperiodic QCSSs  with  a flexible alphabet size  with respect to the newly derived bound  \cite{PS2}.
Besides, in \cite{XLC}, Xiao et al. constructed three classes of asymptotically optimal aperiodic QCSSs with small alphabet sizes.

Since QCSSs have many promising applications in MC-CDMA systems, constructing new QCSSs with large set sizes and low maximum correlation magnitudes has been an interesting but challenging work. This paper focuses on new constructions of both asymptotically optimal periodic and asymptotically optimal aperiodic QCSSs with set size $\Theta(K^2)$ for some block size $K$.
Let $p$ be a prime and $n$ a positive integer. The main contributions of this paper are as follows:
\begin{enumerate}
\item[$\bullet$] Based on the additive characters and permutation polynomials over finite fields, we construct a family of asymptotically optimal periodic QCSSs with parameters $(p^{2n}, p^n-1, p^n-1, p^n+1)$ and alphabet size $p$ (see Theorem \ref{111}) which has larger set size than the known family of $(p^n(p^n-1), p^n-1, p^n-1, p^n+1)$-QCSSs in \cite{XLC}.
\item[$\bullet$] A family of asymptotically optimal aperiodic QCSSs with parameters $((p^n+1)^2, p^n, p^n, p^n)$ and alphabet size $p(p^n+1)$ is constructed by additive characters over finite fields and complex roots of unity (see Theorem \ref{222}). This family of aperiodic QCSSs has much larger set size than the known families of aperiodic QCSSs in \cite{LY5, ZZ1, ZZ2, XLC} when they have the same flock size, length of each constituent sequence and the maximum correlation sum.  
\item[$\bullet$] In Theorem \ref{333}, we construct a family of asymptotically optimal $(p^{2n}+p^n, p^n, p^n, p^n)$ aperiodic QCSSs with alphabet size $p$, which has larger set size than the known family of $(p^{2n}-p^n, p^n, p^n, p^n)$-QCSSs  with alphabet size $p$ in \cite{XLC} and much smaller alphabet size than the known family of $(p^{2n}+p^n, p^n, p^n, p^n)$-QCSSs with alphabet size $p^n$ in \cite{LY5}.
\item[$\bullet$] In Theorem \ref{444}, we present a family of asymptotically optimal $(p^{2n}+p^n-2, p^n, p^n+1, p^n)$ aperiodic QCSSs with alphabet size $p^{n+1}+2p$ which has larger set size than the known family of $(p^{2n}-p^n, p^n, p^n+1, p^n)$ aperiodic QCSSs with alphabet size $p$ in \cite{XLC}. 
\item[$\bullet$] In Theorem \ref{777}, we construct a family of asymptotically optimal $((p^{n}-1)^2, p^n-1, p^n-1, p^n-1)$ aperiodic QCSSs with alphabet size $p^{n}-1$ which has larger set size  than the known family of $(\mathcal{L}\times F(\mathcal{L}), \mathcal{L}, \mathcal{L}, \mathcal{L})$ aperiodic QCSSs with alphabet size $\mathcal{L}$ in \cite{ZZ2} when $\mathcal{L}=p^n-1$, where $F(\mathcal{L})$ $(\leq \mathcal{L})$ is the  maximum number of rows for which an  $\mathcal{L}\times F(\mathcal{L})$ Florentine rectangle exists. For instance, the construction in \cite{ZZ2} can produce a $(4\times26, 26, 26, 26)$-QCSS with alphabet size $26$ for $p=n=3$, while our construction in Theorem \ref{777} can produce a $(26\times26, 26, 26, 26)$-QCSS with alphabet size $26$.
\item[$\bullet$] In Theorem \ref{888}, we obtain a family of asymptotically optimal aperiodic QCSSs with parameters  $(p^{2n}-p^n,p^n,p^n-2,p^n)$. This family of QCSSs have new lengths of the constituent sequences compared with known ones. 
\end{enumerate}
The parameters of the known and proposed periodic and aperiodic QCSSs are listed in Table \ref{tab-1} and Table \ref{tab-2}, respectively.

\begin{table*}[h!]
\begin{center}
\begin{threeparttable}
\caption{The parameters of the known and  proposed asymptotically optimal periodic QCSSs.}\label{tab-1}
\setlength{\tabcolsep}{1mm}{
\begin{tabular}{l|l|l|l|l|l|l}
\hline 
Set size $M$ & Flock size $K$ & Length $N$ & $\vartheta_{\max}$ & Alphabet Size & Constraints& References\\
\hline 
$2^n-1$ & $2^{n-1}-1$ & $2^n$& $(2^n+2^\frac{n}{2})/2$ & $2(2^n-1)$&$n>1$ &\cite{ZL} \\
\hline 
$p$ & $(p-1)/2$ &$p$ & $\leq(p+\sqrt{p})/2$& $p$&$p$ is a prime &\cite{LY1} \\
\hline 
$p^n-2$ & $(p^n-1)/2$ & $p^n-1$ &$\leq (p^n+4 \sqrt{p^n}+3)/2$ &$p^n-1$ & $p$ is an odd prime&\cite{LY2}\\
\hline 
$p^{2n}-2$ & $p^n$&$p^{2n}-1$ &$p^n(p^{2n})+3$ & $p^{2n}-1$&$p$ is a prime&\cite{LY2}\\
\hline 
$p^n-1$ &   $(p^n-1)/2$&$p^n-1$ & $\leq(p^n+\sqrt{p^n})/2$&$p(p^n-1)$& $p$ is an odd prime&\cite{LY3}\\
 \hline 
$p^n-1$ & $p^{n-1}$&$p^n-1$ &$\leq p^{n-\frac{1}{2}}$ & $p(p^{n}-1)$&$p$ is a prime and $n>1$ &\cite{LY3}\\
\hline 
$p^{2n}-1$ &$p^n$ & $p^{2n}-1$& $p^{3n/2}$ &$p(p^{2n}-1)$ & $p$ is a prime&\cite{LY3}\\
\hline 
$2^n-1$& $2^{n-1}-1$& $2^n-1$&$2^{n-1}$ & $2(2^n-1)$&$n>1$ &\cite{LY3}\\
\hline 
$p^n$&$(p^n-1)/2 $&$p^n-1$&$(p^n+1)/2$&$p$& $p$ is an odd prime, $p^n>3$&\cite{LG}\\
\hline 
$p^n$&$(p^n-p^{n-1})/2$&$p^n-1$&$(p^n+p^{n-1})/2$&$p$&$p$  is an odd prime, $p^n>3$&\cite{LG}\\
\hline 
$p^n$&$u$&$p^n-1$&$v$&$p$&$p$ is an odd prime, $n>1$ is odd &\cite{LG}\\
\hline 
$p^{2n}$&$p^n$&$p^n-1$&$p^n$&$p$&$p$ is an odd prime&\cite{ZH1}\\
\hline 
$2^{2n}$&$2^n$&$2^n-1$&$2^n$&$2$&$n\geq1$&\cite{ZH1}\\
\hline 
$p^{2n}-p^n$&$p^n-1$&$p^n-1$&$p^n$&$p(p^n-1)$&$p$ is a prime&\cite{ZH1}\\
\hline 
$p^{2n}-p^n$&$p^n$&$p^n-1$&$p^n$&$p$& $p$ is a prime, $n>1$ &\cite{XLC}\\
\hline 
$p^{2n}-p^n$&$p^n-1$&$p^n-1$&$p^n+1$&$p$& $p$ is a prime, $n>1$ &\cite{XLC}\\
\hline 
$p^{2n}$&$p^n-1$&$p^n-1$&$p^n+1$&$p$&$p$ is a prime, $n>1$&Theorem \ref{111}\\
\hline
\end{tabular}}

\begin{tablenotes}
\footnotesize
\item[]$u=\frac{p^n-p^{n-1}+(-1)^{(p-1)(n+3)/4}(p-1)p^{(n-1)/2}}{2}$, $v=\frac{p^n+p^{n-1}-(-1)^{(p-1)(n+3)/4}(p-1)p^{(n-1)/2}}{2}$.
\end{tablenotes}

\end{threeparttable}
\end{center}

\end{table*}

\begin{table*}[h!]
\begin{center}
\begin{threeparttable}
\caption{The parameters of the known and  proposed  asymptotically optimal aperiodic QCSSs.}\label{tab-2}
\setlength{\tabcolsep}{1mm}{
\begin{tabular}{l|l|l|l|l|l|l}
\hline 
Set size $M$& Flock size $K$& Length $N$& $\theta_{\max}$ & Alphabet Size& Constraints& References\\
\hline 
$p^{2n}+p^n$ & $p^n$ & $p^n$& $p^n$ & $p^n$&$p$ is a prime &\cite{LY5} \\
\hline 
$p^{2n}-p^n$ & $p^n-1$ &$p^n$ & $p^n-1$& $p^{2n}-p^n$&$p$ is a prime, $p^n\geq3$ &\cite{LY5} \\
\hline 
$p^{2n}$ & $p^n$ & $p^n-1$ &$p^n$ &$p^n$ &$p$ is a prime, $p^n\geq3$ &\cite{LY5}\\
\hline 
$\mathcal{L}(p_0-1)$ & $\mathcal{L}$&$\mathcal{L}$ &$\mathcal{L}$ & $\mathcal{L}$&
$\mathcal{L}\geq5 $ is odd, $\mathcal{L}=p_{0}^{e_{0}}p_{1}^{e_{1}}\cdots p_{n-1}^{e_{n-1}}$ &\cite{ZZ1}\\
&&&&& where $p_0\leq p_1\leq\cdots\leq p_{n-1}$ are prime factors of $\mathcal{L}$&\\
\hline 
$\mathcal{L}\times F(\mathcal{L})$ &   $\mathcal{L}$&$\mathcal{L}$ & $\mathcal{L}$&$\mathcal{L}$& $\mathcal{L}\geq 2$ is any integer, $F(\mathcal{L}) (\leq \mathcal{L})$ is the maximum number of rows&\cite{ZZ2}\\
 &&&&&for which an $\mathcal{L}\times F(\mathcal{L})$ Florentine rectangle exists&\\
 \hline 
$p^{n+1}(p-1)$ & $p^{n+1}$&$p^m$ &$p^m$ & $q$& $p$ is a prime and $m$ is a positive integer, $n(\leq m-1)$&\cite{PS2}\\
&&&&& is any non-negative integer, $q$ is a positive multiple of $p$&\\
\hline 
$p^{2n}-1$ &$p^n+1$ & $p^{n}+1$& $p^{n}+1$ &$p^{n}+1$ & $p^n\geq4$ &\cite{WM1}\\
\hline 
$2^{2n}+2^n$& $2^{n}$& $2^n-1$&$2^{n}$ & $2$& $n$ is a positive integer &\cite{WM1}\\
\hline 
$p^{2n}+p^n$&$p^n$&$p^n-1$&$p^n$&$p$& $p$ is a prime &\cite{XLC}\\

\hline 
$p^{2n}-p^n$&$p^n$&$p^n$&$p^n$&$p$& $p^n>3$  &\cite{XLC}\\
\hline 
$p^{2n}-p^n$&$p^n$&$p^n+1$&$p^n$&$p$& $p^n>3$  &\cite{XLC}\\
\hline 
$q(p_0-1)$&$q$&$q-t$&$q$&$\lambda$& $q\geq2$ is any integer,  $p_0$ is the smallest prime factor of $q$, &\cite{ZZ3}\\
&&&&&$t\in\{0, 1\}$, and $q\mid\lambda$&\\
\hline
$q^m(p_0-1)$&$q^m$&$q^m-t$&$q^m$&$\lambda$& $q\geq2$ is any integer, $m\geq1$, $p_0$ is the smallest prime factor of $q$,&\cite{ZZ3}\\
&&&&&$t\in\{0, 1\}$, and $q\mid\lambda$&\\
\hline
$(p^n+1)^2$&$p^n$&$p^n$&$p^n$&$p^{n+1}+p$&$p$ is a prime&Theorem \ref{222}\\
\hline
$p^{2n}+p^n$&$p^n$&$p^n$&$p^n$&$p$&$p$ is a prime&Theorem \ref{333}\\
\hline
$p^{2n}+p^n-2$&$p^n$&$p^n+1$&$p^n$&$p^{n+1}+2p$&$p$ is a prime& Theorem \ref{444}\\
\hline
$(p^n-1)^2$&$p^n-1$&$p^n-1$&$p^n-1$&$p^{n}-1$&$p$ is a prime, $p^n>2$&Theorem \ref{777}\\
\hline
$p^{2n}-p^n$&$p^n$&$p^n-2$&$p^n$&$p^{n+1}-p$&$p$ is a prime, $p^n>2$&Theorem \ref{888}\\
\hline
\end{tabular}}
\end{threeparttable}
\end{center}
\end{table*}

\section{Preliminaries}\label{sec1}
In this section, we recall some basic definitions of QCSSs, additive characters over finite fields and some results which will be used in this paper.
\subsection{Periodic and aperiodic QCSSs}
Let $\textbf{a}=(a_0,a_1,\cdots,a_{N-1})$ and $\textbf{b}=(b_0,b_1,\cdots,b_{N-1})$ be two complex-valued sequences of length $N$. Define the periodic correlation function between them by
$$R_{\textbf{a},\textbf{b}}(\tau)=\sum_{t=0}^{N-1}a_t\overline{b_{t+\tau}}$$
for $0\leq \tau \leq N-1$, where $(t+\tau)$ is calculated modulo $N$ and $\overline{b_{t+\tau}}$ denotes the complex conjugation of $b_{t+\tau}$. Define the aperiodic correlation function between them by
\begin{eqnarray*}
T_{\textbf{a},\textbf{b}}(\tau)=\left\{
\begin{array}{lll}
\sum_{t=0}^{N-\tau-1}a_t\overline{b_{t+\tau}}    &   \mbox{ for }0\leq \tau <N,\\
\sum_{t=0}^{N+\tau-1}a_{t-\tau}\overline{b_{t}}    &   \mbox{ for }-N< \tau <0.
\end{array} \right.
\end{eqnarray*}
Let $\mathcal{C}=\{\mathbf{C}^{0},\mathbf{C}^{1},\cdots,\mathbf{C}^{M-1}\}$ be a set of $M$ complementary sequences, where each $\mathbf{C}^{m}$ is a  two-dimensional matrix with size $K \times N$ given as
\begin{eqnarray*}
\mathbf{C}^{m}=\left[
\begin{array}{cccc}
\mathbf{c}_{0}^{m}\\
\mathbf{c}_{1}^{m}\\
\vdots\\
\mathbf{c}_{K-1}^{m}\\
\end{array}\right]
=
\left[
\begin{array}{cccc}
c_{0,0}^{m} & c_{0,1}^{m}  & \cdots & c_{0,N-1}^{m}\\
c_{1,0}^{m}  & c_{1,1}^{m}  & \cdots & c_{1,N-1}^{m}\\
\vdots   & \vdots  & \ddots & \vdots \\
c_{K-1,0}^{m}  & c_{K-1,1}^{m} & \cdots & c_{K-1,N-1}^{m} \\
\end{array}\right],
\end{eqnarray*}
and $\mathbf{c}_{k}^{m}=(c_{k,0}^{m}, c_{k,1}^{m},\cdots, c_{k,N-1}^{m})$ is the $k$-th constituent sequence of length $N$, $0\leq k \leq K-1$, $0\leq m\leq M-1$.
The periodic correlation function (sum) between two complementary sequences $\mathbf{C}^{{m}_{1}}$ and $\mathbf{C}^{{m}_{2}}$ is defined by 
\begin{eqnarray*}
R_{{{\textbf{C}}^{{m}_{1}}},{{\textbf{C}}^{{m}_{2}}}}(\tau)=\sum_{k=0}^{K-1}R_{{\mathbf{c}_{k}^{{m}_{1}}},{\mathbf{c}_{k}^{{m}_{2}}}}(\tau), ~ 0\leq \tau < N ,
\end{eqnarray*}
where $0\leq m_{1},m_{2}\leq M-1$. The aperiodic correlation function (sum) between $\mathbf{C}^{{m}_{1}}$ and $\mathbf{C}^{{m}_{2}}$ is defined as 
\begin{eqnarray*}
T_{{{\textbf{C}}^{{m}_{1}}},{{\textbf{C}}^{{m}_{2}}}}(\tau)=\sum_{k=0}^{K-1}T_{{\mathbf{c}_{k}^{{m}_{1}}},{\mathbf{c}_{k}^{{m}_{2}}}}(\tau), ~ 0\leq |\tau| < N ,
\end{eqnarray*}
where $0\leq m_{1},m_{2}\leq M-1$. 
The  maximum periodic  auto-correlation magnitude and the maximum periodic  cross-correlation magnitude of $\mathcal{C}$ are  respectively defined by
\begin{eqnarray*}
\vartheta_{a}=\max\{\vert R_{{{\mathbf{C}}^{{m}}},{{\mathbf{C}}^{{m}}}}(\tau)\vert:0\leq m< M, 0<\tau< N\},\\
\vartheta_{c}=\max\{\vert R_{{{\mathbf{C}}^{{m}_{1}}},{{\mathbf{C}}^{{m}_{2}}}}(\tau)\vert:0\leq m_{1}\neq m_{2}< M, 0\leq\tau< N\}.
\end{eqnarray*}
The  maximum aperiodic  auto-correlation magnitude and the maximum aperiodic  cross-correlation magnitude of $\mathcal{C}$ are  respectively defined by
\begin{eqnarray*}
\theta_{a}= \max\{\vert T_{{{\mathbf{C}}^{{m}}},{{\mathbf{C}}^{{m}}}}(\tau)\vert:0\leq m< M, 0<\tau< N\},\\
\theta_{c}=\max\{\vert T_{{{\mathbf{C}}^{{m}_{1}}},{{\mathbf{C}}^{{m}_{2}}}}(\tau)\vert:0\leq m_{1}\neq m_{2}< M, 0\leq\tau< N\}.
\end{eqnarray*}
The maximum periodic (resp. aperiodic) correlation magnitude, also called periodic (resp. aperiodic) tolerance, is a performance measure
of the sequence set $\mathcal{C}$ in practical applications, which is defined as
\begin{eqnarray*}
 \vartheta_{\max} = \max\{\vartheta_{c}, \vartheta_{a}\}~ (\text{resp. }  \theta_{\max} = \max\{\theta_{c}, \theta_{a}\}).
\end{eqnarray*}
If $0<\vartheta_{\max}\ll KN$ (resp. $0<\theta_{\max}\ll KN$) and $M>K>1$, then $\mathcal{C}$ is called a periodic (resp. aperiodic) $(M, K, N, \vartheta_{\max}$) (resp. $(M, K, N, \theta_{\max}$)) quasi-complementary sequence set (QCSS).
\subsection{The correlation lower bounds of periodic and aperiodic QCSSs}
In \cite{ZL}, Liu et al. presented a lower bound for  a periodic $(M, K ,N, \vartheta_{\max})$-QCSS as follows:
\begin{eqnarray}\label{zl1}
\vartheta_{\max}\geq\vartheta_{\text{opt}}=KN\sqrt{\frac{M/K-1}{MN-1}}.
\end{eqnarray}

In \cite{Welch}, Welch established a lower bound for an aperiodic $(M, K ,N, \theta_{\max})$-QCSS as follows:
\begin{eqnarray}\label{wlech1}
\theta_{\max}\geq \theta_{\text{opt}}=KN\sqrt{\frac{M/K-1}{M(2N-1)-1}}.
\end{eqnarray}
In 2014, Liu et al. gave the following tighter lower bound of $\theta_{\max}$ for an aperiodic $(M, K ,N, \theta_{\max})$-QCSS.
\begin{lemma}\cite{ZL1}
For an aperiodic $(M, K ,N, \theta_{\max})$-QCSS with $M\geq3K$, $K\geq2$ and $N\geq2$, we have 
\begin{eqnarray}\label{zl2}
\theta_{\max}\geq \theta_{\textup{opt}}=\sqrt{KN\left(1-2\sqrt{\frac{K}{3M}}\right)}.
\end{eqnarray}
\end{lemma}

If $\rho_0:=\frac{\vartheta_{\max}}{\vartheta_{\text{opt}}}=1$ (resp.~ $\rho_1 := \frac{\theta_{\max}}{\theta_{\text{opt}}}=1$), then the periodic (resp. an aperiodic) QCSS is said to be optimal. If $1<\rho_0 \leq 2$ (resp. $1<\rho_1\leq2$), then the periodic (resp. an aperiodic) QCSS is said to near-optimal.
\subsection{Additive and multiplicative characters of finite fields}
 Let $q=p^n$, where $p$ is a prime and $n$ is a positive integer. Denote by $\zeta_N$ the primitive $N$-th root of complex unity for a positive integer $N$.
Let $\gf_q$ denote the finite field with $q$ elements. Define the trace function from $\gf_q$ to $\gf_p$ by
$$\tr_{q/p}(x)=x+x^p+\cdots+x^{p^{n-1}},\ x\in \gf_q.$$
An \emph{additive character}  of $\gf_q$ is defined as the homomorphic function $\chi_a$ from $\gf_q$ to $\mathbb{C}^*$, which can be given by
$$\chi_a(x)=\zeta_{p}^{\tr_{q/p}(ax)},\ x\in \gf_q.$$
By definition, $\chi_a(x)=\chi_1(ax)$. $\chi_0$  and $\chi_1$ are called the trivial and canonical additive characters of $\gf_q$, respectively.
The complex conjugate $\overline{\chi_a}$ of $\chi_a$ is defined by $\overline{\chi_a}(x)=\overline{\chi_a(x)}=\chi_a(-x)$ for $x\in \gf_q$.
 In \cite{Lidl}, the orthogonality relation of additive characters is given by
\begin{eqnarray*}
\sum_{x\in \gf_q}\chi_1(ax)=\left\{
\begin{array}{ll}
q  &   \mbox{if $a=0$},\\
0    &   \mbox{if $a\in \gf_q^*$}.
\end{array} \right.
\end{eqnarray*}

Let $\alpha$ be a primitive element of $\gf_q$.
A  \emph{multiplicative character} of $\gf_q$ is defined as   a homomorphism $\varphi$ from $\gf_q^*$ to $\mathbb{C}^*$ such that  $\varphi(xy)=\varphi(x)\varphi(y)$ for any $x,y\in \gf_q^*$.
Then 
$$\varphi_j(\alpha^k)=\zeta_{q-1}^{jk},\ k=0,1,\cdots,q-2,$$
for each $j=0,1,\cdots,q-2$, defines a multiplicative character. In particular, $\varphi_0$  is called the trivial  multiplicative character. For odd $q$, $\eta:=\varphi_{\frac{q-1}{2}}$  is referred to as the quadratic multiplicative character of $\gf_q$. 
The complex conjugate $\overline{\varphi}$ of a multiplicative character $\varphi$ is defined by $\overline{\varphi}(x)=\overline{\varphi(x)}=\varphi(x^{-1})$ for $x\in \gf_q^*$.
From \cite{Lidl}, the orthogonality relation of multiplicative characters is given by 
\begin{eqnarray*}
\sum_{x\in \gf_q^*}\varphi_j(x)=\left\{
\begin{array}{ll}
q-1  &   \mbox{if $j=0$},\\
0    &   \mbox{if $1\leq j \leq q-2$}.
\end{array} \right.
\end{eqnarray*}

\subsection{The $m$-sequence}
For a positive integer $r$, let $\alpha$ be a primitive element of $\gf_{q^r}$.
 The well-known $q$-ary $m$-sequence of period $q^r-1$ is defined by
 $$\mathbf{s}=(s(0), s(1), \cdots,s(q^r-2)),$$
 where $s(j)=\tr_{q^r/q}(\alpha^j)=\sum_{i=0}^{r-1}\alpha^{q^ij}$ with $0\leq j\leq q^r-2$.

The following lemma is vital to calculate the aperiodic correlation magnitude of some QCSSs in this paper.
\begin{lemma}\cite{MKS}\label{lem-2}
Let $\mathbf{s}$ be the $m$-sequence defined above.
Then every segment of $\frac{q^r-1}{q-1}$ consecutive symbols from $\mathbf{s}$ contains exactly $\frac{q^{r-1}-1}{q-1}$ zeros.
\end{lemma}

\section{A new construction of periodic QCSSs}\label{sec1+}
In this section, we present a new construction of asymptotically optimal periodic QCSSs. Hereafter, we denote by $[t_1,t_2]$ the set of integers from $t_1$ to $t_2$.

Let $d_0, d_1,\cdots, d_{q-2}$ denote all the elements of $\gf_q^*$. Let $q=p^{n}$, where $p$ is a prime and $n$ is a positive integer. Let $\alpha$ be a primitive element of $\gf_q$. 
Let $\chi_{1}$ be the canonical additive character of $\gf_q$ and $f(x)\in \gf_q[x]$. For $(a,b)\in \gf_q \times \gf_q$, we define a constituent sequence $\mathbf{c}_{l}^{a,b}$ of period $q-1$ as
\begin{eqnarray*}\label{eqna-1}
\mathbf{c}_{l}^{a,b}=\left(\mathbf{c}_{l}^{a,b}(t)\right)_{t=0}^{q-2}, \mbox{  }\mathbf{c}_{l}^{a,b}(t)=\chi_{1}\left(d_{l}\left(f(\alpha^{t})+a\right)+b\alpha^{t}\right) \\ \mbox{ for } 0\leq l\leq q-2.
\end{eqnarray*}
Then we obtain a two-dimensional $(q-1)\times(q-1)$ matrix 
$\mathbf{C}^{a,b}:=\left[\mathbf{c}_{0}^{a,b}, \mathbf{c}_{1}^{a,b},\cdots, \mathbf{c}_{q-2}^{a,b}\right]^{T}$.
Denote by
\begin{eqnarray}\label{eq-cc1}
\mathcal{C}=\left\{\mathbf{C}^{a,b}: a\in\gf_q, b\in \gf_q\right\}
\end{eqnarray}
which is a complementary sequence set.

\begin{theorem}\label{111}
Let $q=p^n$, where $p$ is a prime and $n>1$ is a positive integer. Let $f(x)\in \gf_q[x]$ such that $f(zx)-f(x)$ permutes $\gf_q$ for every $z\in \gf_q\backslash \{1\}$. Let $\mathcal{C}$ be the complementary sequence set
defined in Equation (\ref{eq-cc1}). Then $\mathcal{C}$ is a periodic $(q^2, q-1, q-1, q+1)$-QCSS with alphabet size $p$ which is asymptotically optimal with respect to the lower bound in (\ref{zl1}).
\end{theorem}

\begin{IEEEproof}
For any $(a_{1},b_{1})\in \gf_q\times \gf_q$, $(a_{2},b_{2})\in \gf_q\times \gf_q$ and $\tau\in[0,q-2]$, we have
\begin{eqnarray*}
&&R_{\mathbf{C}^{a_{1},b_{1}},\mathbf{C}^{a_{2},b_{2}}}(\tau)=\sum_{l=0}^{q-2}R_{\mathbf{c}_{l}^{a_{1},b_{1}},\mathbf{c}_{l}^{a_{2},b_{2}}}(\tau)
\\
&=&\sum_{l=0}^{q-2}\sum_{t=0}^{q-2}\chi_{1}\left(d_{l}(f(\alpha^{t})+a_1)+b_1\alpha^{t})\right)
\\
&&\overline{\chi_{1}}\left(d_{l}(f(\alpha^{t+\tau})+a_2)+b_2\alpha^{t+\tau})\right)
\\
&=&\sum_{l=0}^{q-2}\sum_{t=0}^{q-2}\chi_{1}((f(\alpha^{t})-f(\alpha^{t+\tau}))d_{l}+(a_{1}-a_{2})d_{l}+\\
&&(b_{1}-b_{2}\alpha^{\tau})\alpha^{t}).
\end{eqnarray*}
Then it is divided into the following two cases to determine the value distribution of $R_{\mathbf{C}^{a_{1},b_{1}},\mathbf{C}^{a_{2},b_{2}}}(\tau)$.

Case 1: If $\tau=0$, then 
\begin{eqnarray*}
\nonumber &
&R_{\mathbf{C}^{a_{1},b_{1}},\mathbf{C}^{a_{2},b_{2}}}(\tau)\\
\nonumber 
&=&\sum_{l=0}^{q-2}\sum_{t=0}^{q-2}\chi_{1}\left((a_{1}-a_{2})d_l+(b_1-b_2)\alpha^{t}\right)\\
&=&\sum_{x\in \gf_q^*}\chi_{1}\left((a_1-a_2)x\right)\sum_{y\in \gf_q^*}\chi_{1}\left((b_1-b_2)y\right)\\
&=&
\left\{
\begin{array}{ll}
-(q-1) & \text{if} ~ a_1=a_2, b_1\neq b_2,\\
-(q-1) & \text{if} ~ a_1\neq a_2,b_1=b_2,\\
1 &  \text{if} ~ a_1\neq a_2, b_1\neq b_2,
\end{array}\right.
\end{eqnarray*}
where the third equation holds owing to the orthogonality relation of additive characters.

Case 2: If $\tau\neq0$, then 
\begin{eqnarray*}
&&R_{\mathbf{C}^{a_{1},b_{1}},\mathbf{C}^{a_{2},b_{2}}}(\tau)\\
&=&\sum_{l=0}^{q-2}\sum_{t=0}^{q-2}\chi_{1}((f(\alpha^{t})-f(\alpha^{t+\tau}))d_l+(a_{1}-a_{2})d_l+\\
&&(b_{1}-b_{2}\alpha^{\tau})\alpha^{t})\\
&=&\sum_{t=0}^{q-2}\chi_{1}((b_1-b_2\alpha^{\tau})\alpha^{t})\sum_{l=0}^{q-2}\chi_{1}((f(\alpha^{t})-f(\alpha^{t+\tau}))d_l+\\
&&(a_1-a_2)d_l)\\
&=&\sum_{y\in\gf_q^{*}}\chi_{1}((b_1-b_2\alpha^{\tau})y)\sum_{x\in\gf_q^*}\chi_{1}((f(y)-f(y\alpha^{\tau}))x+\\
&&(a_1-a_2)x).
\end{eqnarray*}
Since $f(y)-f(y\alpha^{\tau})=0$ for $y=0$ and  $f(y)-f(y\alpha^{\tau})$ permutes $\gf_q$ for $\tau\neq 0$, we deduce $f(y)-f(y\alpha^{\tau})\neq 0$ for any $y\in \gf_q^*$. 
Then we compute $R_{\mathbf{C}^{a_{1},b_{1}},\mathbf{C}^{a_{2},b_{2}}}(\tau)$ in the following four subcases.

Subcase 2.1: If $a_1-a_2=0$, $b_1-b_2\alpha^\tau=0$, then 
\begin{eqnarray*}
\nonumber &
&R_{\mathbf{C}^{a_{1},b_{1}},\mathbf{C}^{a_{2},b_{2}}}(\tau)\\
\nonumber
&=&\sum_{y\in\gf_q^{*}}\sum_{x\in\gf_q^*}\chi_{1}(\left(f(y)-f(y\alpha^{\tau})\right)x)=-(q-1),
\end{eqnarray*}
by the orthogonality relation of additive characters.

Subcase 2.2: If $a_1-a_2=0$, $b_1-b_2\alpha^\tau\neq0$, then 
\begin{eqnarray*}
\nonumber &
&R_{\mathbf{C}^{a_{1},b_{1}},\mathbf{C}^{a_{2},b_{2}}}(\tau)\\
\nonumber
&=&\sum_{y\in\gf_q^{*}}\chi_{1}((b_1-b_2\alpha^\tau)y)\sum_{x\in\gf_q^*}\chi_{1}(\left(f(y)-f(y\alpha^{\tau})\right)x)\\
&=&1,
\end{eqnarray*}
by the orthogonality relation of additive characters.

Subcase 2.3: If $a_1-a_2\neq0$, $b_1-b_2\alpha^\tau=0$, then 
\begin{eqnarray*}
\nonumber &
&R_{\mathbf{C}^{a_{1},b_{1}},\mathbf{C}^{a_{2},b_{2}}}(\tau)\\
\nonumber
&=&\sum_{x\in\gf_q^{*}}\sum_{y\in\gf_q^{*}}\chi_{1}(\left(f(y)-f(y\alpha^{\tau})\right)x+(a_1-a_2)x)\\
&=&\sum_{x\in\gf_q^{*}}\chi_{1}((a_1-a_2)x)\sum_{y\in\gf_q^*}\chi_{1}\left(\left(f(y)-f(y\alpha^{\tau})\right)x\right)\\
&=&\sum_{x\in\gf_q^{*}}\chi_{1}((a_1-a_2)x)\sum_{y'\in\gf_q^*}\chi_{1}\left(y'x\right)\\
&=&1,
\end{eqnarray*}
by the orthogonality relation of additive characters and the condition that $f(y)-f(y\alpha^{\tau})$ permutes $\gf_q^*$.

Subcase 2.4: If $a_1-a_2\neq0$, $b_1-b_2\alpha^\tau\neq0$, then
\begin{eqnarray*}
\begin{split}
&R_{\mathbf{C}^{a_{1},b_{1}},\mathbf{C}^{a_{2},b_{2}}}(\tau)=\\
&\sum_{y\in\gf_q^{*}}\chi_{1}((b_1-b_2\alpha^{\tau})y)\sum_{x\in\gf_q^*}\chi_{1}((f(y)-f(y\alpha^{\tau})+a_1-a_2)x).
\end{split}
\end{eqnarray*}
Since $f(y)-f(y\alpha^{\tau})$ permutes $\gf_q^{*}$, we know that $\mathcal{Z}(y)=f(y)-f(yz)+a_1-a_2$ has exactly one zero in $\gf_q^*$.
Denote by $y_1$ the unique zero of $\mathcal{Z}(y)$ in $\gf_q^*$. Then
\begin{eqnarray*}
\begin{split}
&R_{\mathbf{C}^{a_{1},b_{1}},\mathbf{C}^{a_{2},b_{2}}}(\tau)\\
&=(q-1)\chi_{1}((b_1-b_2\alpha^{\tau})y_1)+\sum_{\substack{y\in\gf_q^{*}\\ y \neq y_1}}\chi_{1}((b_1-b_2\alpha^{\tau})y)\\
&\sum_{x\in\gf_q^*}\chi_{1}(x(f(y)-f(y\alpha^{\tau})+a_1-a_2))\\
&=(q-1)\chi_{1}((b_1-b_2\alpha^{\tau})y_1)+(-1)\sum_{\substack{y\in\gf_q^{*}\\ y \neq y_1}}\chi_{1}((b_1-b_2\alpha^{\tau})y)\\
&=q\chi_{1}((b_1-b_2\alpha^{\tau})y_1)+1.
\end{split}
\end{eqnarray*}
This implies that $|R_{\mathbf{C}^{a_{1},b_{1}},\mathbf{C}^{a_{2},b_{2}}}(\tau)|\leq q+1$.
It is clear that there exists $(b_1,b_2)\in \gf_q\times \gf_q$ such that $b_1-b_2\alpha^\tau\neq0$ and $\tr_{q/p}((b_1-b_2\alpha^{\tau})y_1)=0$ as $q-1>q-p^{n-1}$ for $n>1$.
 Then there exists $(b_1,b_2)\in \gf_q\times \gf_q$ for $b_1-b_2\alpha^\tau\neq0$ such that $R_{\mathbf{C}^{a_{1},b_{1}},\mathbf{C}^{a_{2},b_{2}}}=q+1$.

Based on the discussions above, we drive that the maximum periodic correlation magnitude of $\mathcal{C}$ is $q+1$.

Next, we verify that  the constructed periodic QCSS is asymptotically optimal. Since  $\mathcal{C}$ is a periodic $(q^2, q-1, q-1, q+1)$-QCSS, according to the bound of (\ref{zl1}), we have 
\begin{eqnarray*}
\vartheta_{\textup{opt}}=(q-1)(q-1)\sqrt{\frac{\frac{q^2}{q-1}-1}{q^2(q-1)-1}}.
\end{eqnarray*}
It is easy to see that 
\begin{eqnarray*}
\lim_{q\rightarrow+\infty}\rho_0=\lim_{q\rightarrow+\infty}\frac{q+1}{(q-1)(q-1)\sqrt{\frac{\frac{q^2}{q-1}-1}{q^2(q-1)-1}}}=1.
\end{eqnarray*}
Then the desired conclusion follows.
\end{IEEEproof}

\begin{remark}
We remark that  there exist many $f(x)\in \gf_q[x]$ such that $f(zx)-f(x)$ permutes $\gf_q$ for every $z\in \gf_q\backslash \{1\}$.
In \cite{Cao1}, Cao et al. listed some families of such polynomials including the followings:
\begin{enumerate}
\item[$(1)$] $f(x)=ax^{d}$, where $a\in \gf_q^*$ and $\gcd(d,q-1)=1$;
\item[$(2)$] $f(x)=\sum_{i=0}^{m-1}a_{i}x^{p^{i}}\in\gf_q[x]$, where $q=p^m$ and $f(x)$ is a permutation polynomial over $\gf_q$.
\end{enumerate}
 \end{remark}

\begin{example}\label{example2}
Let $p=3$, $n=2$ and $f(x)=x$. By Magma program, we verify that $\mathcal{C}$ is a $(81, 8, 8, 10)$-QCSS with alphabet $\{e^{2\pi\sqrt{-1}i/3}: i\in[0,2]\}$. By Python program, we show the auto-correlation magnitude distributions of $\mathbf{C}^{1,\alpha}$, $\mathbf{C}^{1,\alpha^2}$ and $\mathbf{C}^{\alpha,\alpha^2}$in Fig. \ref{fig1} and the cross-correlation magnitude distributions of $\mathbf{C}^{1,\alpha}$, $\mathbf{C}^{1,\alpha^2}$ and $\mathbf{C}^{\alpha,\alpha^2}$in Fig. \ref{fig2}, where the red line stands for the maximum correlation magnitude.

\begin{figure}[H]
    \centering
		\includegraphics[width=0.3\linewidth]{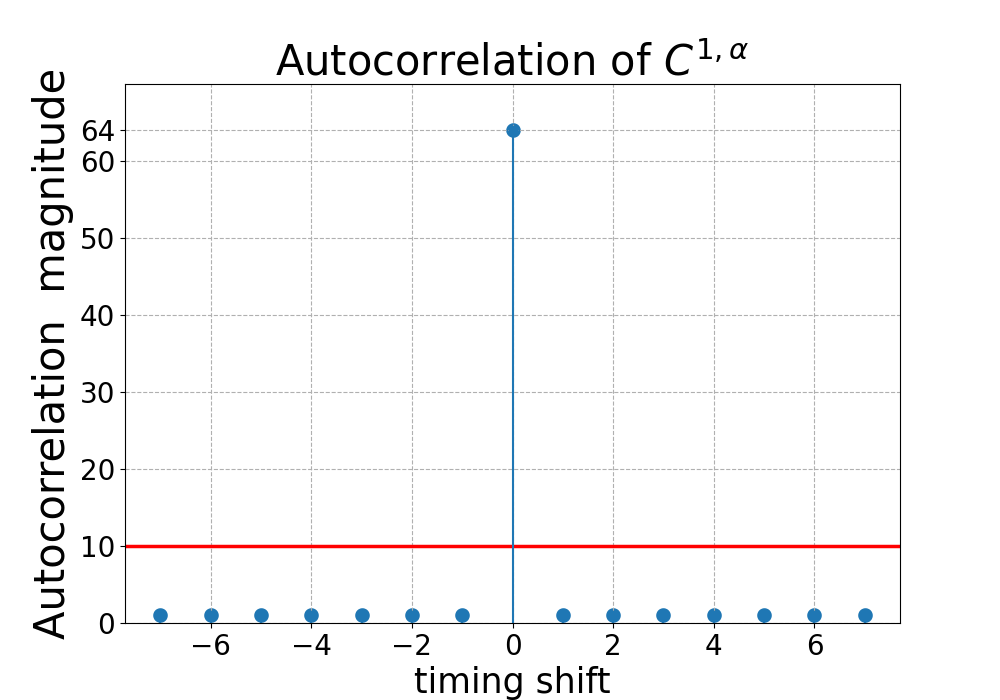}
    \hfill
		\includegraphics[width=0.3\linewidth]{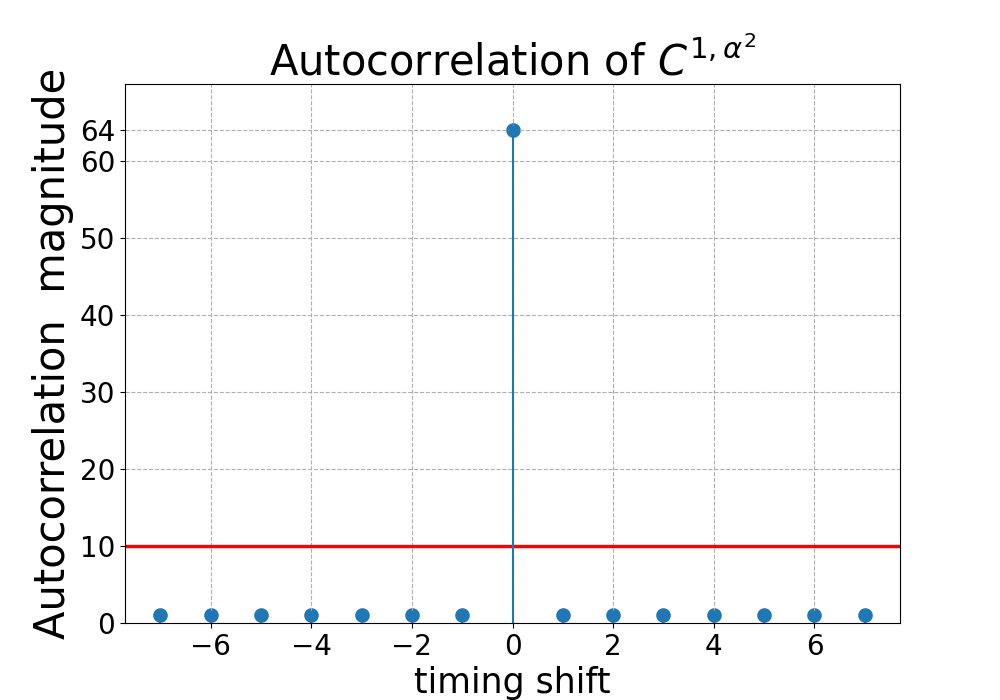}
    \hfill
		\includegraphics[width=0.3\linewidth]{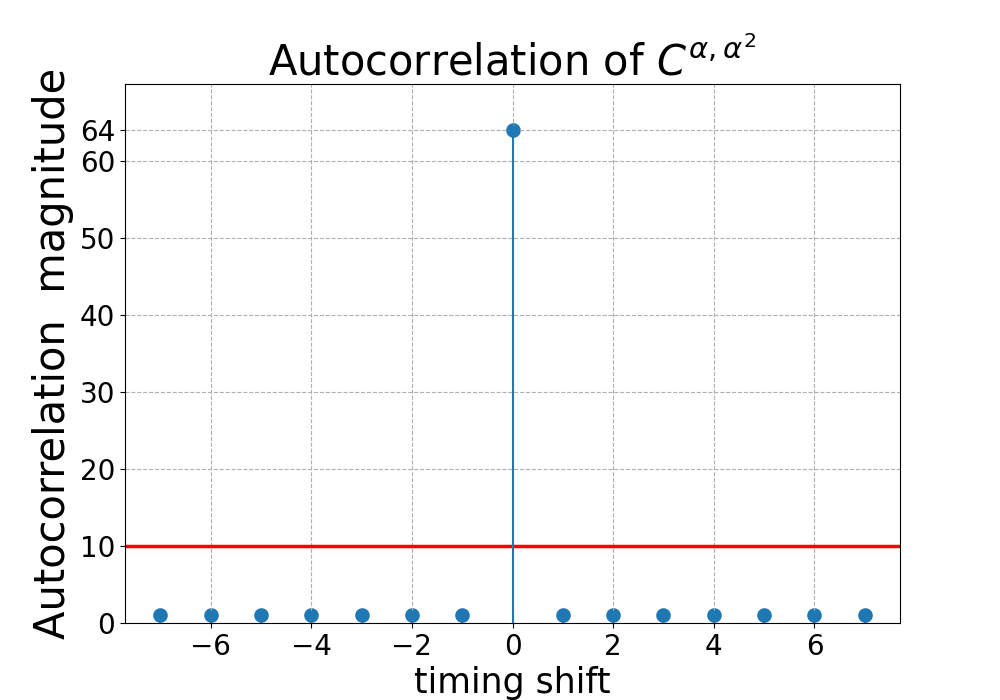}
    \caption{The auto-correlation magnitude distributions of $\mathbf{C}^{1,\alpha}$, $\mathbf{C}^{1,\alpha^2}$ and $\mathbf{C}^{\alpha,\alpha^2}$.} 
    \label{fig1} 
\end{figure}

\begin{figure}[H]
    \centering
		\includegraphics[width=0.3\linewidth]{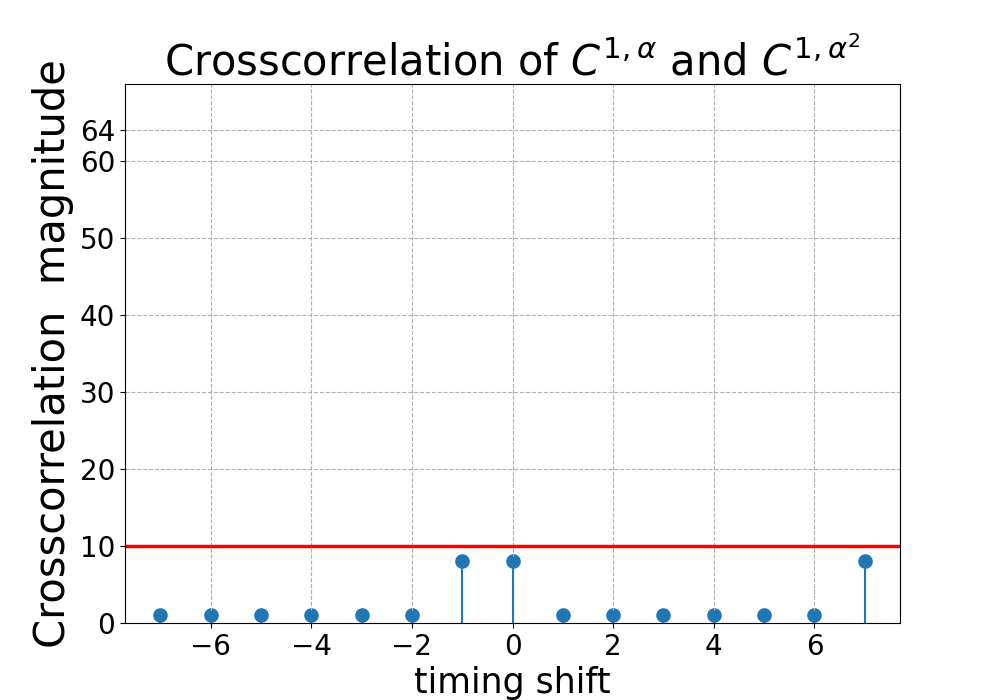}
    \hfill
		\includegraphics[width=0.3\linewidth]{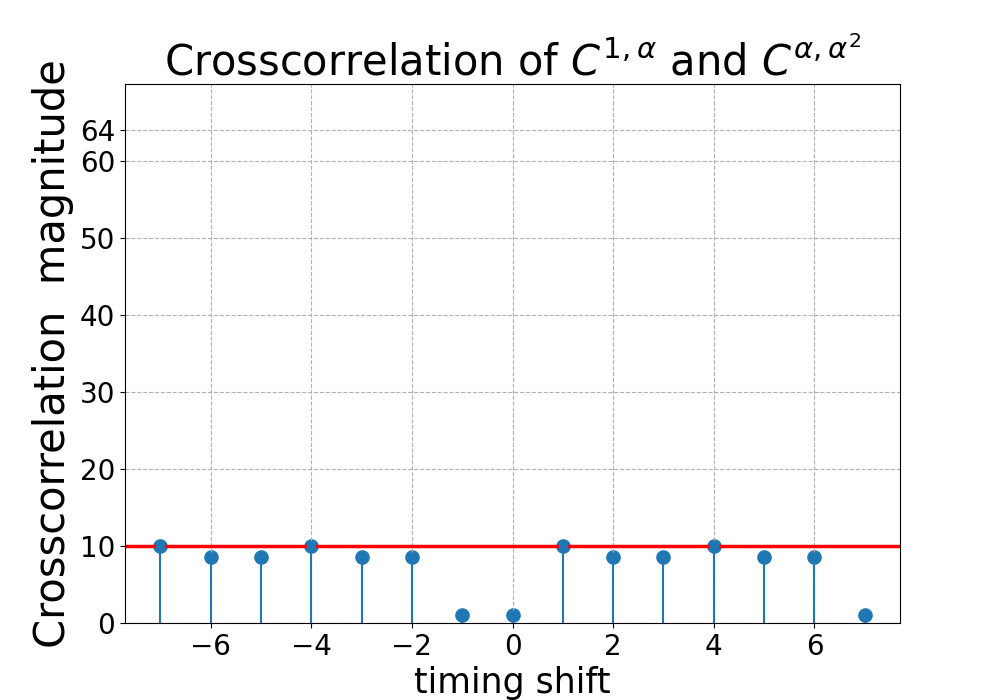}
    \hfill
		\includegraphics[width=0.3\linewidth]{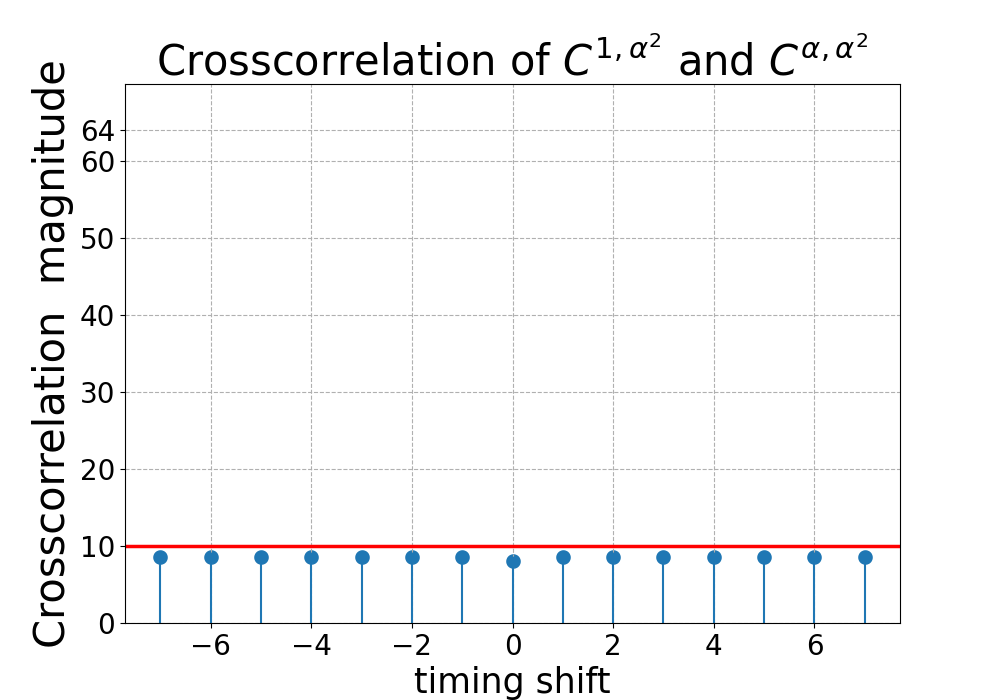}
    \caption{The cross-correlation magnitude distributions of $\mathbf{C}^{1,\alpha}$, $\mathbf{C}^{1,\alpha^2}$ and $\mathbf{C}^{\alpha,\alpha^2}$.} 
    \label{fig2} 
\end{figure}
 
\end{example}

\section{Five new constructions of aperiodic QCSSs}

The objective of this section is to present five new constructions of aperiodic QCSSs with large set sizes which are asymptotically optimal with respect to the lower bound in (\ref{zl2}).
In this section, we let $\alpha$ be a primitive element of $\gf_{q}$,
 $\beta$ be a primitive element of $\gf_{q^2}$ and  $d_0, d_1,\cdots, d_{q-2} ,d_{q-1}$ be all the elements of $\gf_q$, where $q=p^n$ for a prime $p$ and a positive integer $n$. Let $\mathbb{Z}_q$ denote the residue classes ring of integers modulo $q$ and $\Bbb Z$ be the set of all integers. 

\subsection{The first construction of aperiodic QCSSs}
Let $a,b\in \Bbb Z$. Construct a set 
\begin{eqnarray}\label{eqn-2}
\mathcal{A}=\{\mathbf{A}^{a,b}: 0\leq a\leq q, 0\leq b\leq q\},
\end{eqnarray}
 where $\mathbf{A}^{a,b}=\left[\mathbf{s}_{0}^{a,b},\mathbf{s}_{1}^{a,b},\cdots,\mathbf{s}_{q-1}^{a,b}\right]^{T}$
consists of $K=q$ sequences of length $N=q$ defined by
\begin{eqnarray*}
&\mathbf{s}_{k}^{a,b}=(\mathbf{s}_{k}^{a,b}(t))_{t=0}^{q-1}, ~~ \\ &\mathbf{s}_{k}^{a,b}(t)=\chi_1\left(d_k\tr_{q^2/q}(\beta^{a(q-1)+t})\right)\zeta_{q+1}^{bt}, ~ 0\leq k\leq q-1.
\end{eqnarray*}
 It is obvious that the alphabet size of $\mathcal{A}$ is $p(q+1)$.
\begin{theorem}\label{222}
Let $q=p^n$ for a prime $p$ and a positive integer $n$. Then the set $\mathcal{A}$ defined in (\ref{eqn-2}) is an aperiodic $((q+1)^2, q, q, q)$-QCSS which is asymptotically optimal with respect to the bound in (\ref{zl2}).
\end{theorem}

\begin{IEEEproof}
Let $0\leq a_1,a_2\leq q, 0\leq b_1,b_2\leq q$.
For any two complementary sequences $\mathbf{A}^{a_{1},b_{1}}$, $\mathbf{A}^{a_{2},b_{2}}$ in $\mathcal{A}$ and $\tau\in[0,q-1]$, we have
\begin{eqnarray}\label{eqn-T}
\nonumber& &T_{\mathbf{A}^{a_{1},b_{1}},\mathbf{A}^{a_{2},b_{2}}}(\tau)=\sum_{k=0}^{q-1}T_{\mathbf{s}_{k}^{a_{1},b_{1}},\mathbf{s}_{k}^{a_{2},b_{2}}}(\tau)
\\
\nonumber &=&\sum_{k=0}^{q-1}\sum_{t=0}^{q-1-\tau}\chi_1\left(d_k\tr_{q^2/q}(\beta^{a_1(q-1)+t})\right)\zeta_{q+1}^{b_1t}\\
\nonumber &&\overline{\chi}_1\left(d_k\tr_{q^2/q}(\beta^{a_2(q-1)+t+\tau})\right)\zeta_{q+1}^{-b_2(t+\tau)}\\
\nonumber &=&\zeta_{q+1}^{-b_2\tau}\sum_{t=0}^{q-1-\tau}\zeta_{q+1}^{(b_1-b_2)t}\sum_{k=0}^{q-1}\chi_1\Big(d_k\big(\tr_{q^2/q}(\beta^{a_1(q-1)+t})\\
&&-\tr_{q^2/q}(\beta^{a_2(q-1)+t+\tau})\big)\Big).
\end{eqnarray}
We divide into the following cases to determine the value distribution of $T_{\mathbf{A}^{a_{1},b_{1}},\mathbf{A}^{a_{2},b_{2}}}(\tau)$.

{Case 1}: If $\tau=0$, $a_1=a_2$ and $b_1\neq b_2$, then  
\begin{eqnarray*}
\begin{split}
&T_{\mathbf{A}^{a_{1},b_{1}},\mathbf{A}^{a_{2},b_{2}}}(\tau)= \sum_{k=0}^{q-1}\sum_{t=0}^{q-1}\zeta_{q+1}^{(b_1-b_2)t}\\
&=q\left(\sum_{t=0}^{q-1}\zeta_{q+1}^{(b_1-b_2)t}+\zeta_{q+1}^{(b_1-b_2)q}-\zeta_{q+1}^{(b_1-b_2)q}\right)\\
&=q\left(\sum_{t=0}^{q}\zeta_{q+1}^{(b_1-b_2)t}-\zeta_{q+1}^{(b_1-b_2)q}\right)
=-q\zeta_{q+1}^{(b_1-b_2)q},
\end{split}
\end{eqnarray*}
where $0<|b_1-b_2|\leq q$ and
$$\sum_{t=0}^{q}\zeta_{q+1}^{(b_1-b_2)t}=\frac{\zeta_{q+1}^{0}-\zeta_{q+1}^{(b_1-b_2)(q+1)}}{1-\zeta_{q+1}^{b_1-b_2}}=0.$$

{Case 2}: If $\tau\neq0$ and $a_1=a_2$, then
\begin{eqnarray*}
&
 &T_{\mathbf{A}^{a_{1},b_{1}},\mathbf{A}^{a_{2},b_{2}}}(\tau)\\
&=&\zeta_{q+1}^{-b_2\tau}\sum_{t=0}^{q-1-\tau}\zeta_{q+1}^{(b_1-b_2)t}\sum_{k=0}^{q-1}\chi_1\Big(d_k\\
&&\tr_{q^2/q}\left(\beta^{a_1(q-1)+t}(1-\beta^{\tau})\right)\Big)
\end{eqnarray*}
by Equation (\ref{eqn-T}).
Since $1\leq \tau \leq q-1$, we deduce that $\tr_{q^2/q}\left(\beta^{a_1(q-1)+t}(1-\beta^{\tau})\right)=0$ with the variable $t$ has at most one solution for $0\leq t \leq q-1-\tau$ by Lemma \ref{lem-2}.

{Subcase 2.1}: If $\tr_{q^2/q}\left(\beta^{a_1(q-1)+t}(1-\beta^{\tau})\right)=0$ has one solution $t_0 \in [0,q-1-\tau]$, then 
\begin{eqnarray*}
& &T_{\mathbf{A}^{a_{1},b_{1}},\mathbf{A}^{a_{2},b_{2}}}(\tau)\\
&=&\zeta_{q+1}^{-b_2\tau}q\zeta_{q+1}^{(b_1-b_2)t_0}+\zeta_{q-1}^{-b_2\tau}\sum_{\substack{t=0\\ t \neq t_0}}^{q-1-\tau}\zeta_{q+1}^{(b_1-b_2)t}\\
&&\sum_{k=0}^{q-1}\chi_1\left(d_k\tr_{q^2/q}\left(\beta^{a_1(q-1)+t}(1-\beta^{\tau})\right)\right)\\
&=&\zeta_{q+1}^{-b_2\tau}q\zeta_{q+1}^{(b_1-b_2)t_0}+\zeta_{q-1}^{-b_2\tau}\sum_{\substack{t=0\\ t \neq t_0}}^{q-1-\tau}\zeta_{q+1}^{(b_1-b_2)t}\\
&&\sum_{y\in\gf_q}\chi_1\left(y\tr_{q^2/q}\left(\beta^{a_1(q-1)+t}(1-\beta^{\tau})\right)\right)\\
&=&q\zeta_{q+1}^{(b_1-b_2)t_0-b_2\tau},
\end{eqnarray*}
where the third equation holds owing to the orthogonality relation of additive characters.

{Subcase 2.2}: If $\tr_{q^2/q}\left(\beta^{a_1(q-1)+t}(1-\beta^{\tau})\right)=0$ has no solution for $0\leq t \leq q-1-\tau$, then we have
 \begin{eqnarray*}
\begin{split}
&T_{\mathbf{A}^{a_{1},b_{1}},\mathbf{A}^{a_{2},b_{2}}}(\tau)\\
&=\zeta_{q+1}^{-b_2\tau}\sum_{t=0}^{q-1-\tau}\zeta_{q+1}^{(b_1-b_2)t}\\
&\sum_{y\in\gf_q}\chi_{1}\left(y\tr_{q^2/q}\left(\beta^{a_1(q-1)+t}(1-\beta^{\tau})\right)\right)=0
\end{split}
\end{eqnarray*}
by the orthogonality relation of additive characters.

{Case 3}: If $\tau\in[0,q-1]$ and $a_1\neq a_2$, then we have
\begin{eqnarray*}
\nonumber &
&T_{\mathbf{A}^{a_{1},b_{1}},\mathbf{A}^{a_{2},b_{2}}}(\tau)\\
&=&\zeta_{q+1}^{-b_2\tau}\sum_{t=0}^{q-1-\tau}\zeta_{q+1}^{(b_1-b_2)t}\sum_{k=0}^{q-1}\chi_1\Big(d_k\big(\tr_{q^2/q}
(\beta^{a_1(q-1)+t}\\
&&(1-\beta^{(a_2-a_1)(q-1)+\tau}))\big)\Big)
\end{eqnarray*}
by Equation (\ref{eqn-T}).

{Subcase 3.1}:
If $\tau=q-1$ and $a_2-a_1=q$, then $1-\beta^{(a_2-a_1)(q-1)+\tau}=0$ and $T_{\mathbf{A}^{a_{1},b_{1}},\mathbf{A}^{a_{2},b_{2}}}(\tau)=q\zeta_{q+1}^{-b_2(q-1)}$.

{Subcase 3.2}: If $(\tau,a_2-a_1)\neq (q-1,q)$, then it is easy to verify that $1-\beta^{(a_2-a_1)(q-1)+\tau}\neq0$. According to Lemma \ref{lem-2}, we drive that $\tr_{q^2/q}\left(\beta^{a_1(q-1)+t}(1-\beta^{(a_1-a_2)(q-1)+\tau})\right)=0 $ has at most one solution for $0\leq t \leq q-1-\tau$.  Similarly to the {Case 2} above, we deduce
\begin{eqnarray*}
& &| T_{\mathbf{A}^{a_{1},b_{1}},\mathbf{A}^{a_{2},b_{2}}}(\tau)| =\\
& & \left\{
\begin{array}{ll}
q & \text{if} ~ \tr_{q^2/q}(\beta^{a_1(q-1)+t}(1-\beta^{(a_1-a_2)(q-1)+\tau}))=0 \\& \text{~has one solution for }0\leq t \leq q-1-\tau,\\
0 & \text{if} ~ \tr_{q^2/q}(\beta^{a_1(q-1)+t}(1-\beta^{(a_1-a_2)(q-1)+\tau}))=0 \\& \text{~has no solution for }0\leq t \leq q-1-\tau.
\end{array}\right.
\end{eqnarray*}

Summarizing the above three cases, we obtain that the maximum aperiodic correlation magnitude of $\mathcal{A}$ is $q$.
Finally, we prove that  $\mathcal{A}$ is asymptotically optimal according to
the bound in (\ref{zl2}). Since $\mathcal{A}$ is an aperiodic $((q+1)^2, q, q, q)$-QCSS, according to the lower bound in (\ref{zl2}), we have 
\begin{eqnarray*}
\theta_{\textup{opt}}=\sqrt{q^2\left(1-2\sqrt{\frac{q}{3(q+1)^2}}\right)}.
\end{eqnarray*}
It is easy to see that 
\begin{eqnarray*}
\lim_{q\rightarrow+\infty}\rho_1
=\lim_{q\rightarrow+\infty}\frac{q}{\sqrt{q^2\left(1-2\sqrt{\frac{q}{3(q+1)^2}}\right)}}=1.
\end{eqnarray*}
Then the proof of this theorem is completed.
\end{IEEEproof}

\begin{example}
Let $p=2$ and $n=3$. 
By Magma program, we verify that the set $\mathcal{A}$ constructed in Theorem \ref{222} is an aperiodic $(81, 8, 8, 8)$-QCSS with alphabet $\{e^{2\pi\sqrt{-1}i/18}: i\in[0,17]\}$.
By Python program, we show the auto-correlation and cross-correlation magnitude distributions of $\mathbf{A}^{1,1}$, $\mathbf{A}^{1,2}$ and $\mathbf{A}^{2,1}$in Fig. \ref{fig3} and Fig. \ref{fig4}, respectively, where the red line stands for the maximum correlation magnitude. 

\begin{figure}[H]
    \centering
		{\includegraphics[width=0.3\linewidth]{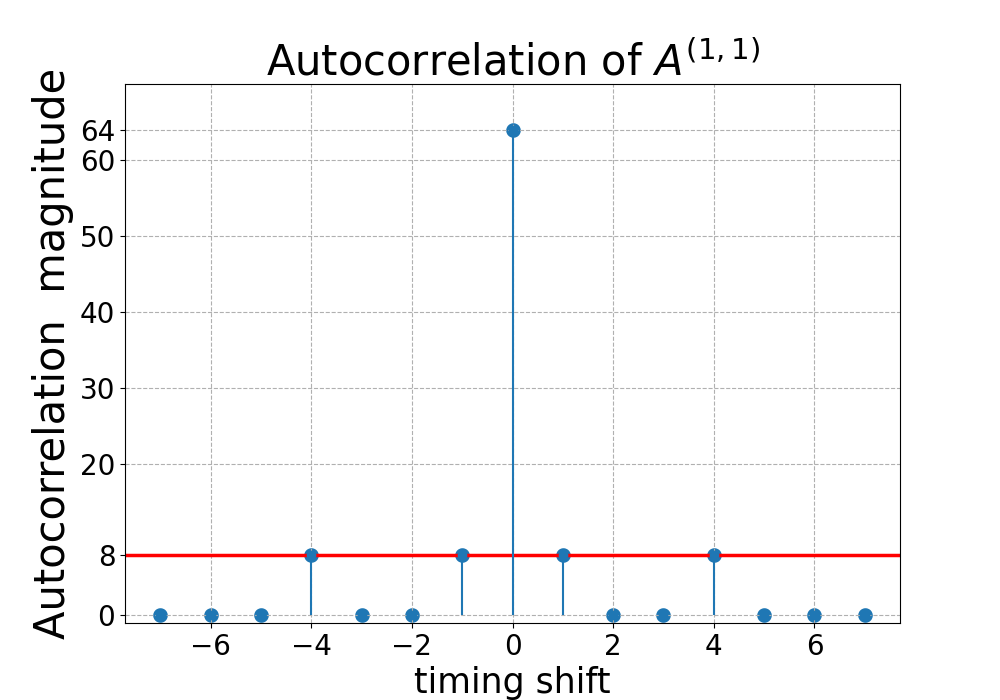}}
    \hfill
		{\includegraphics[width=0.3\linewidth]{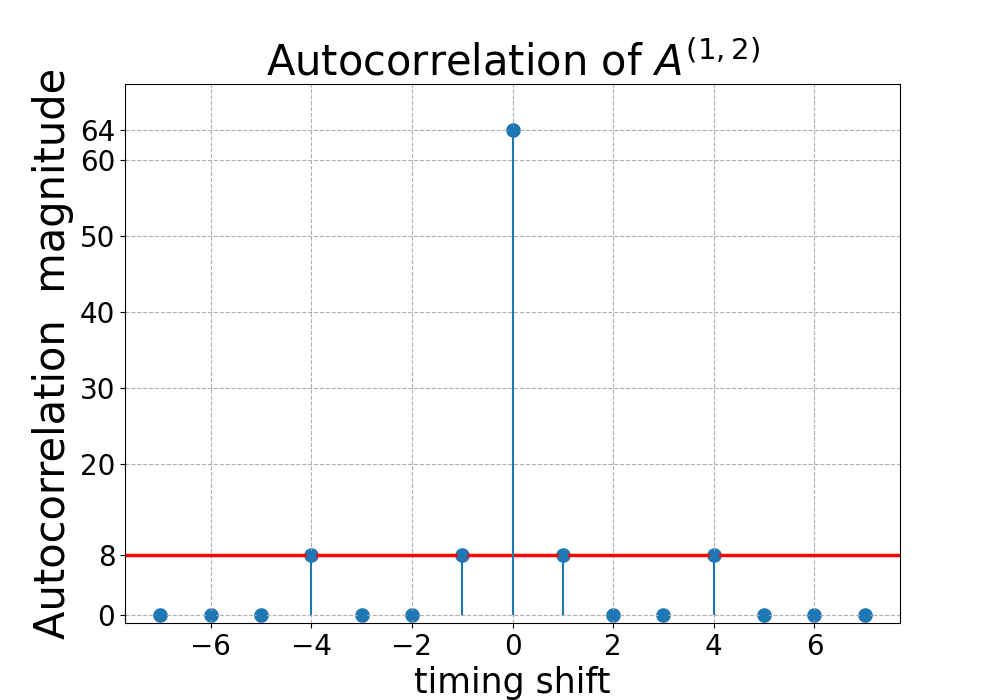}}
    \hfill
		{\includegraphics[width=0.3\linewidth]{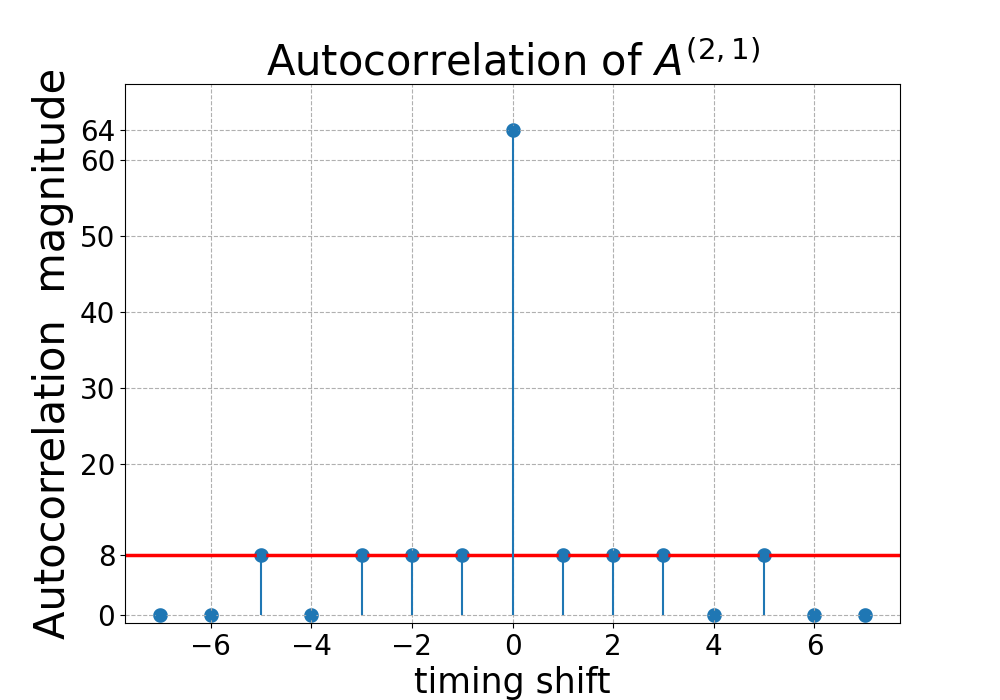}}
    \caption{The auto-correlation magnitude distributions of $\mathbf{A}^{1,1}$, $\mathbf{A}^{1,2}$ and $\mathbf{A}^{2,1}$.} 
    \label{fig3} 
\end{figure}
\begin{figure}[H]
    \centering
		{\includegraphics[width=0.3\linewidth]{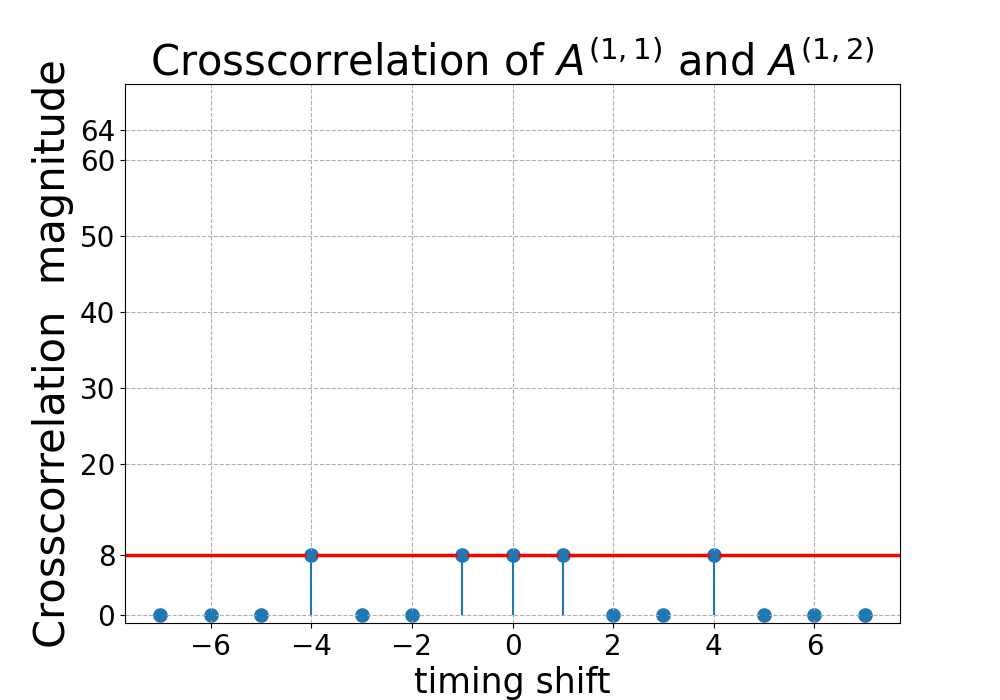}}
    \hfill
		{\includegraphics[width=0.3\linewidth]{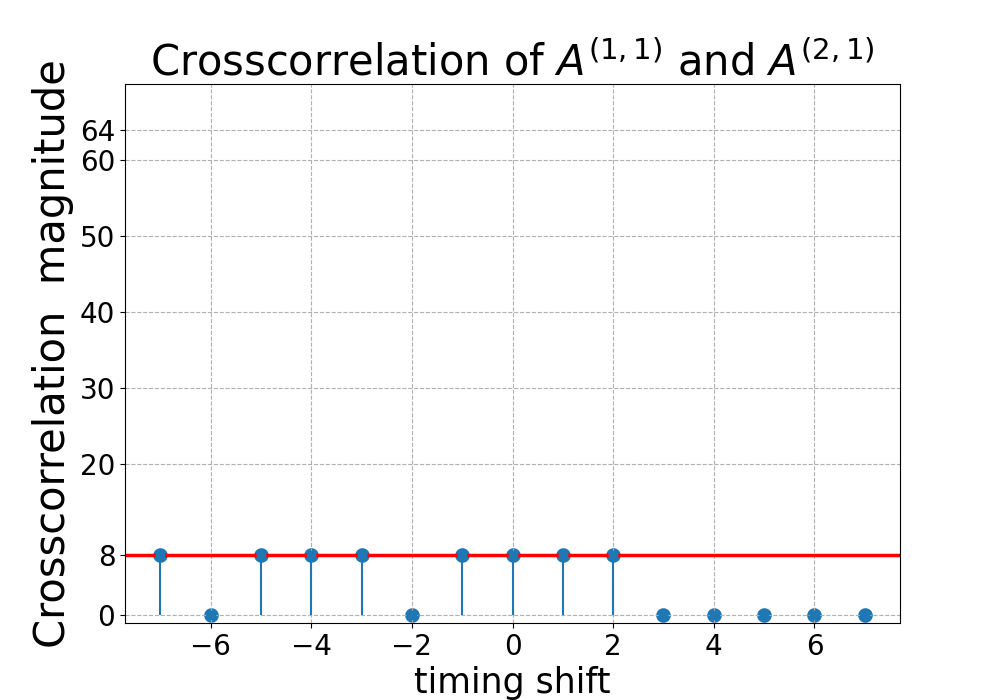}}
    \hfill
		{\includegraphics[width=0.3\linewidth]{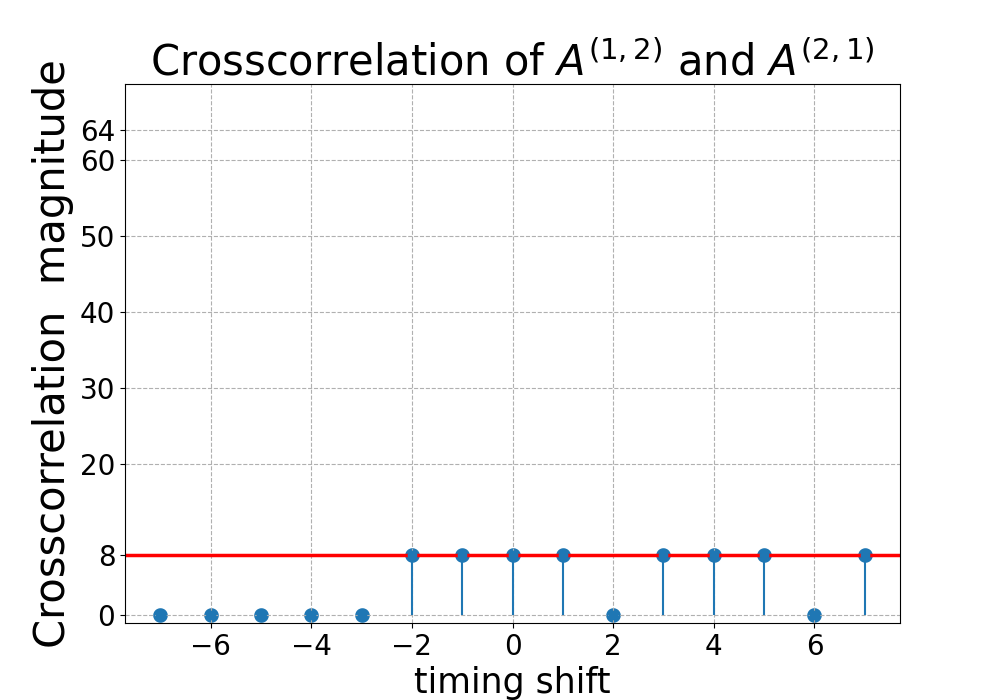}}
    \caption{The cross-correlation magnitude distributions of $\mathbf{A}^{1,1}$, $\mathbf{A}^{1,2}$ and $\mathbf{A}^{2,1}$.} 
    \label{fig4} 
\end{figure}
\end{example}

\subsection{The second construction of aperiodic QCSSs}
Define a function $\sigma(x):\mathbb{Z}_{q}\rightarrow \mathbb{Z}_2$, where
\begin{eqnarray*}
\sigma(x)=\left\{
\begin{array}{ll}
1 & \text{if} ~ 0\leq x  \leq q-2,\\
0 & \text{if} ~ x = q-1.
\end{array}\right.
\end{eqnarray*}
Let $a\in \Bbb Z$. Construct a set 
\begin{eqnarray}\label{eqn-3}
\mathcal{B}=\{\mathbf{B}^{a,b}: 0\leq a\leq q,  b\in \gf_q\}
\end{eqnarray}
containing $M=q(q+1)$ complementary sequences, where $\mathbf{B}^{a,b}=\left[\mathbf{s}_{0}^{a,b},\mathbf{s}_{1}^{a,b},\cdots,\mathbf{s}_{q-1}^{a,b}\right]^{T}$
consists of $K=q$ sequences of length $N=q$ defined by
\begin{eqnarray*}
&\mathbf{s}_{k}^{a,b}=(\mathbf{s}_{k}^{a,b}(t))_{t=0}^{q-1}, \\~~ &\mathbf{s}_{k}^{a,b}(t)=\chi_1\left(d_k\tr_{q^2/q}(\beta^{a(q-1)+t})\right)\chi_b\left(\beta^{(q+1)t}\sigma(t)\right), ~\\& 0\leq k\leq q-1.
\end{eqnarray*}
 It is obvious that the alphabet size of $\mathcal{B}$ is $p$.
\begin{theorem}\label{333}
Let $q=p^n$ for a prime $p$ and a positive integer $n$. Then the set $\mathcal{B}$ defined in (\ref{eqn-3}) is an aperiodic $(q(q+1), q, q, q)$-QCSS with alphabet size $p$ which is asymptotically optimal with respect to the bound in (\ref{zl2}).
\end{theorem}
\begin{IEEEproof}
Let $0\leq a_1,a_2\leq q,  b_1,b_2\in \gf_q$.
For any two complementary sequences $\mathbf{B}^{a_{1},b_{1}}$, $\mathbf{B}^{a_{2},b_{2}}$ in $\mathcal{B}$ and $\tau\in[0,q-1]$, we have
\begin{eqnarray}\label{eqn-T-2}
\nonumber & &T_{\mathbf{B}^{a_{1},b_{1}},\mathbf{B}^{a_{2},b_{2}}}(\tau)\\
\nonumber &=&
\sum_{k=0}^{q-1}T_{\mathbf{s}_{k}^{a_{1},b_{1}},\mathbf{s}_{k}^{a_{2},b_{2}}}(\tau)
\\
\nonumber &=&\sum_{k=0}^{q-1}\sum_{t=0}^{q-1-\tau}\chi_1\left(d_k\tr_{q^2/q}(\beta^{a_1(q-1)+t})\right)\\
\nonumber & &\cdot \chi_{b_1}\left(\beta^{(q+1)t}\sigma(t)\right) \overline{\chi}_1\left(d_k\tr_{q^2/q}(\beta^{a_2(q-1)+t+\tau})\right)\\
\nonumber & &\cdot \overline{\chi}_{b_2}\left(\beta^{(q+1)(t+\tau)}\sigma(t+\tau)\right)\\
\nonumber &=&\sum_{t=0}^{q-1-\tau}\chi_1\left(\beta^{(q+1)t}\left(b_1\sigma(t)-b_2\beta^{(q+1)\tau}\sigma(t+\tau)\right)\right)\\
 & &\sum_{k=0}^{q-1}\chi_1\left(d_k\tr_{q^2/q}(\beta^{a_1(q-1)+t}-\beta^{a_2(q-1)+t+\tau})\right). 
\end{eqnarray}
Then we divide into the following cases to determine the value distribution of $T_{\mathbf{B}^{a_{1},b_{1}},\mathbf{B}^{a_{2},b_{2}}}(\tau)$.

{Case 1}: If $\tau=0$, $a_1=a_2$ and $b_1\neq b_2$, then  
\begin{eqnarray*}
\begin{split}
&T_{\mathbf{B}^{a_{1},b_{1}},\mathbf{B}^{a_{2},b_{2}}}(\tau)=\sum_{k=0}^{q-1}\sum_{t=0}^{q-1}\chi_1\left(\beta^{(q+1)t}\sigma(t)(b_1-b_2)\right)\\
&=\sum_{k=0}^{q-1}\sum_{x\in\gf_q}\chi_1(x(b_1-b_2))=0
\end{split}
\end{eqnarray*}
by the definition of $\sigma(t)$ and the orthogonality relation of additive characters.

{Case 2}: If $\tau\neq0$ and $a_1= a_2$, then  
\begin{eqnarray*}
\nonumber &
&T_{\mathbf{B}^{a_{1},b_{1}},\mathbf{B}^{a_{2},b_{2}}}(\tau)\\
&=&\sum_{t=0}^{q-1-\tau}\chi_1\left(\beta^{(q+1)t}\left(b_1\sigma(t)-b_2\beta^{(q+1)\tau}\sigma(t+\tau)\right)\right)
\\&&\sum_{k=0}^{q-1}\chi_1\left(d_k\left(\tr_{q^2/q}\left(\beta^{a_1(q-1)+t}(1-\beta^\tau)\right)\right)\right)
\end{eqnarray*}
by Equation (\ref{eqn-T-2}).
By Lemma \ref{lem-2}, we deduce that $\tr_{q^2/q}\left(\beta^{a_1(q-1)+t}(1-\beta^\tau)\right)=0$ with variable $t$ has at most one solution if $0\leq t  \leq q-1-\tau$.

{Subcase 2.1}:
If $\tr_{q^2/q}\left(\beta^{a_1(q-1)+t}(1-\beta^\tau)\right)\neq 0$ for $0\leq t \leq q-1-\tau$, then 
\begin{eqnarray*}
\nonumber &
&T_{\mathbf{B}^{a_{1},b_{1}},\mathbf{B}^{a_{2},b_{2}}}(\tau)\\
&=&\sum_{t=0}^{q-1-\tau}\chi_1\left(\beta^{(q+1)t}\left(b_1\sigma(t)-b_2\beta^{(q+1)\tau}\sigma(t+\tau)\right)\right)
\\&&\sum_{y\in\gf_q}\chi_1\left(y\left(\tr_{q^2/q}\left(\beta^{a_1(q-1)+t}(1-\beta^\tau)\right)\right)\right)\\
&=&0
\end{eqnarray*}
by the orthogonality relation of additive characters.

{Subcase 2.2}:
If there exists a unique solution $0\leq t_0 \leq q-1-\tau$ of $\tr_{q^2/q}\left(\beta^{a_1(q-1)+t_0}(1-\beta^\tau)\right)= 0$, then 
\begin{eqnarray*}
\nonumber &
&T_{\mathbf{B}^{a_{1},b_{1}},\mathbf{B}^{a_{2},b_{2}}}(\tau)\\
&=&\sum_{t=0}^{q-1-\tau}\chi_1\left(\beta^{(q+1)t}\left(b_1\sigma(t)-b_2\beta^{(q+1)\tau}\sigma(t+\tau)\right)\right)
\\&&\sum_{y\in\gf_q}\chi_1\left(y\left(\tr_{q^2/q}\left(\beta^{a_1(q-1)+t}(1-\beta^\tau)\right)\right)\right)\\
&=&q\chi_1\left(\beta^{(q+1)t_0}\left(b_1\sigma(t_0)-b_2\beta^{(q+1)\tau}\sigma(t_0+\tau)\right)\right)+\\
&&\sum_{\substack{t=0\\ 
t \ne  t_0}}^{q-1-\tau}\chi_1\left(\beta^{(q+1)t}\left(b_1\sigma(t)-b_2\beta^{(q+1)\tau}\sigma(t+\tau)\right)\right)
\\&&\sum_{y\in\gf_q}\chi_1\left(y\left(\tr_{q^2/q}\left(\beta^{a_1(q-1)+t}(1-\beta^\tau)\right)\right)\right)\\
&=&q\chi_1\left(\beta^{(q+1)t_0}\left(b_1\sigma(t_0)-b_2\beta^{(q+1)\tau}\sigma(t_0+\tau)\right)\right)
\end{eqnarray*}
by the orthogonality relation of additive characters.

{Case 3}: If $\tau\in[0,q-1]$ and $a_1\neq a_2$, then 
\begin{eqnarray*}
\begin{split}
&T_{\mathbf{B}^{a_{1},b_{1}},\mathbf{B}^{a_{2},b_{2}}}(\tau)\\
&=\sum_{t=0}^{q-1-\tau}\chi_1\left(\beta^{(q+1)t}\left(b_1\sigma(t)-b_2\beta^{(q+1)\tau}\sigma(t+\tau)\right)\right)
\\&\sum_{k=0}^{q-1}\chi_1\left(d_k\left(\tr_{q^2/q}\left(\beta^{a_1(q-1)+t}(1-\beta^{(a_2-a_1)(q-1)+\tau})\right)\right)\right)\\
\end{split}
\end{eqnarray*}
by Equation (\ref{eqn-T-2}).

{Subcase 3.1}: If $\tau=q-1$ and $a_2-a_1=q$, then $1-\beta^{(a_2-a_1)(q-1)+\tau}=0$ and 
$T_{\mathbf{B}^{a_{1},b_{1}},\mathbf{B}^{a_{2},b_{2}}}(\tau)=q\chi_1(b_1)$.

{Subcase 3.2}: If $(\tau,a_2-a_1) \neq (q-1,q)$, then $1-\beta^{(a_2-a_1)(q-1)+\tau}\neq0$. It is obvious that $$\tr_{q^2/q}\left(\beta^{a_1(q-1)+t}(1-\beta^{(a_1-a_2)(q-1)+\tau})\right)=0 $$ has at most one solution for $0\leq t \leq q-1-\tau$ by Lemma \ref{lem-2}. 
Similarly to the {Case 2} above, we have 
\begin{eqnarray*}
&&\vert T_{\mathbf{B}^{a_{1},b_{1}},\mathbf{B}^{a_{2},b_{2}}}(\tau)\vert\\
& =&\left\{
\begin{array}{ll}
q & \text{if} ~ \tr_{q^2/q}(\beta^{a_1(q-1)+t}(1-\beta^{(a_1-a_2)(q-1)+\tau}))=0\\ &\text{~has one solution for }0\leq t \leq q-1-\tau,\\
0 & \text{Otherwise}.
\end{array}\right.
\end{eqnarray*}

By the discussions above, we drive that the maximum aperiodic correlation magnitude $\theta_{\max}$ of $\mathcal{B}$ is $q$.

In the following, we verify that the obtained aperiodic QCSS is asymptotically optimal according to
the bound in (\ref{zl2}). Since $\mathcal{B}$ is an aperiodic $(q(q+1), q, q, q)$-QCSS, according to the lower bound in (\ref{zl2}), we have 
\begin{eqnarray*}
\theta_{\textup{opt}}=\sqrt{q^2\left(1-2\sqrt{\frac{q}{3q(q+1)}}\right)}.
\end{eqnarray*}
It is easy to see that 
\begin{eqnarray*}
\lim_{q\rightarrow+\infty}\rho_1=\lim_{q\rightarrow+\infty}\frac{\theta_{\max}}{\theta_{\textup{opt}}}=\lim_{q\rightarrow+\infty}\frac{q}{\sqrt{q^2(1-2\sqrt{\frac{q}{3q(q+1)}})}}=1.
\end{eqnarray*}
The desired conclusion follows. 
\end{IEEEproof}

\begin{example}
Let $p=3$ and $n=2$. 
Then the parameters of the QCSS $\mathcal{B}$ constructed in Theorem \ref{333} are $(90, 9, 9, 9)$ and its alphabet is given by $\{e^{2\pi\sqrt{-1}i/3}: i\in[0,2]\}$, which is verified by Magma program.
By Python program, we respectively show the auto-correlation and cross-correlation magnitude distributions of $\mathbf{B}^{1,\alpha}$, $\mathbf{B}^{1,\alpha^2}$ and $\mathbf{B}^{2,\alpha}$ in Fig. \ref{fig5} and Fig. \ref{fig6}, where the red line stands for the maximum correlation magnitude. 

\begin{figure}[H]
    \centering
		{\includegraphics[width=0.3\linewidth]{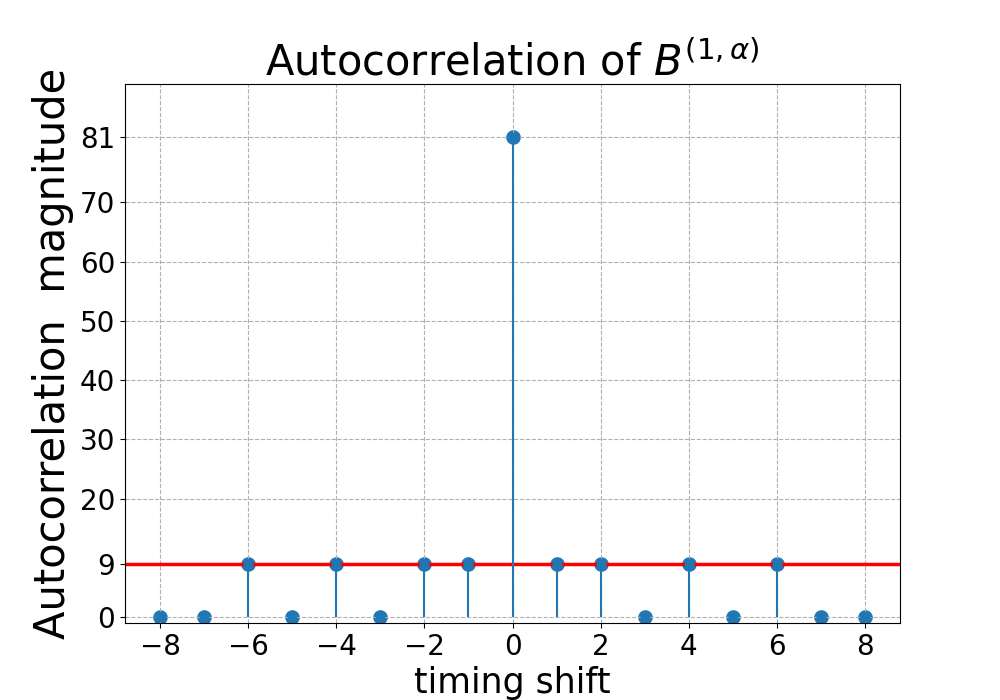}}
    \hfill
		{\includegraphics[width=0.3\linewidth]{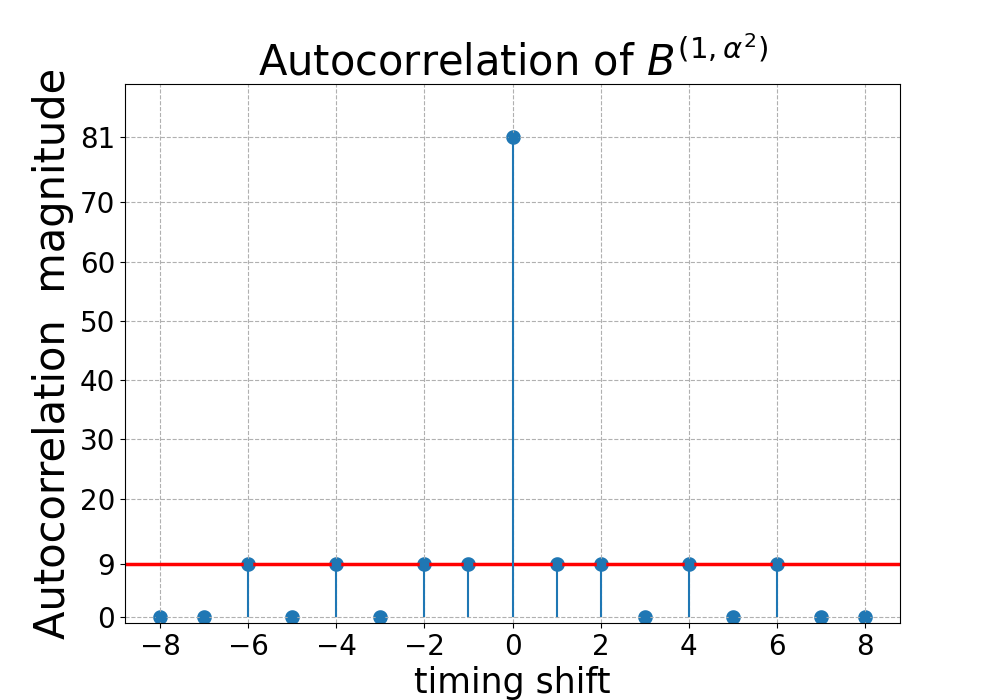}}
    \hfill
		{\includegraphics[width=0.3\linewidth]{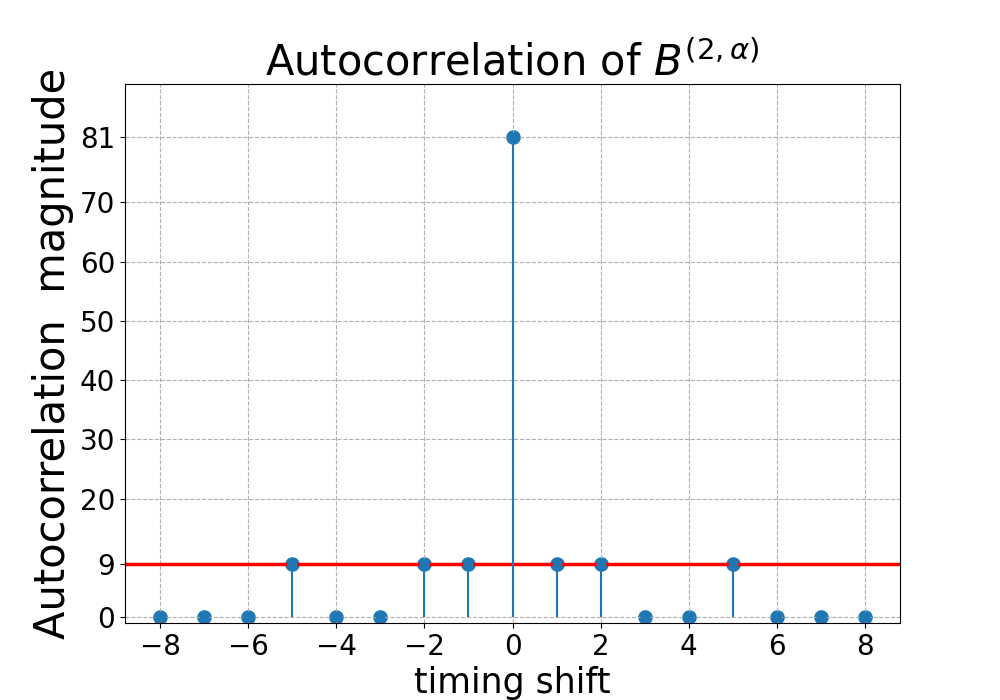}}
    \caption{The auto-correlation magnitude distributions of $\mathbf{B}^{1,\alpha}$,$\mathbf{B}^{1,\alpha^2}$ and $\mathbf{B}^{2,\alpha}$.} 
    \label{fig5} 
\end{figure}
\begin{figure}[H]
    \centering
		{\includegraphics[width=0.3\linewidth]{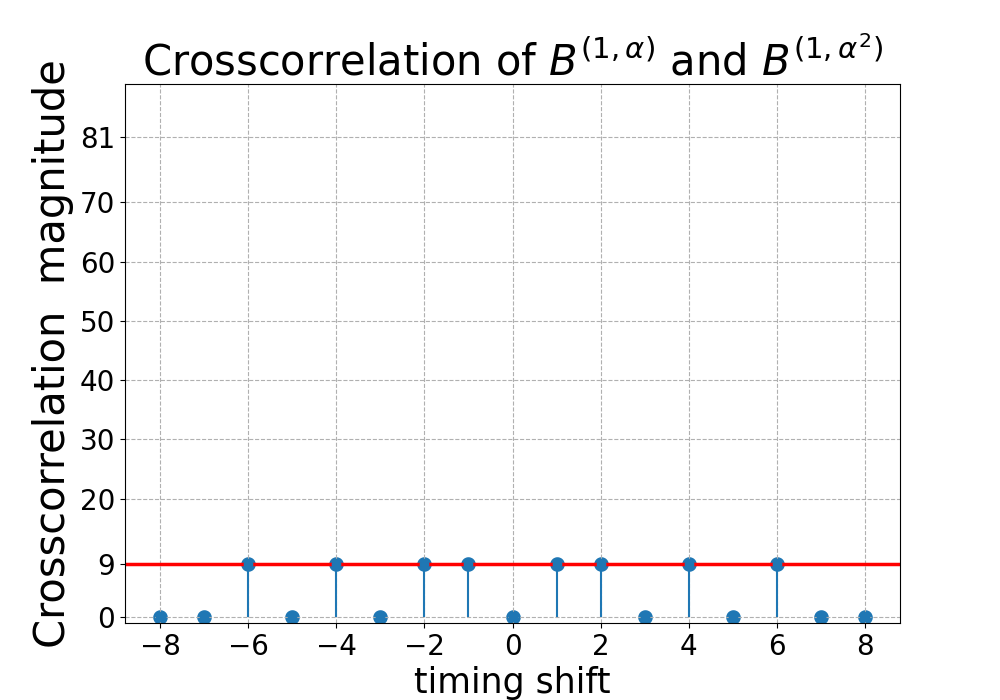}}
    \hfill
		{\includegraphics[width=0.3\linewidth]{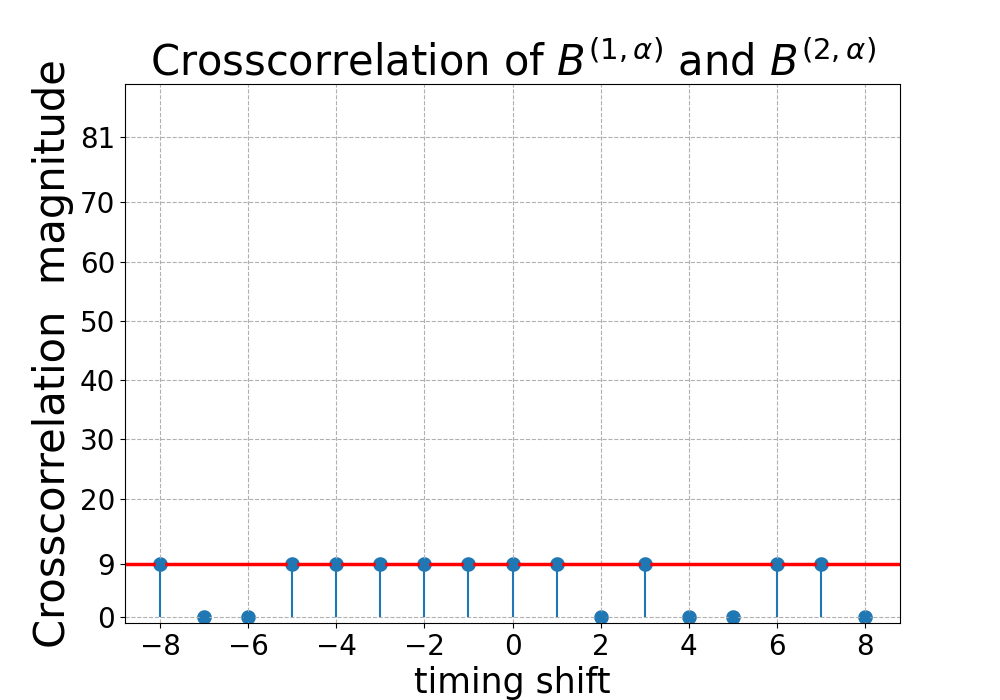}}
    \hfill
		{\includegraphics[width=0.3\linewidth]{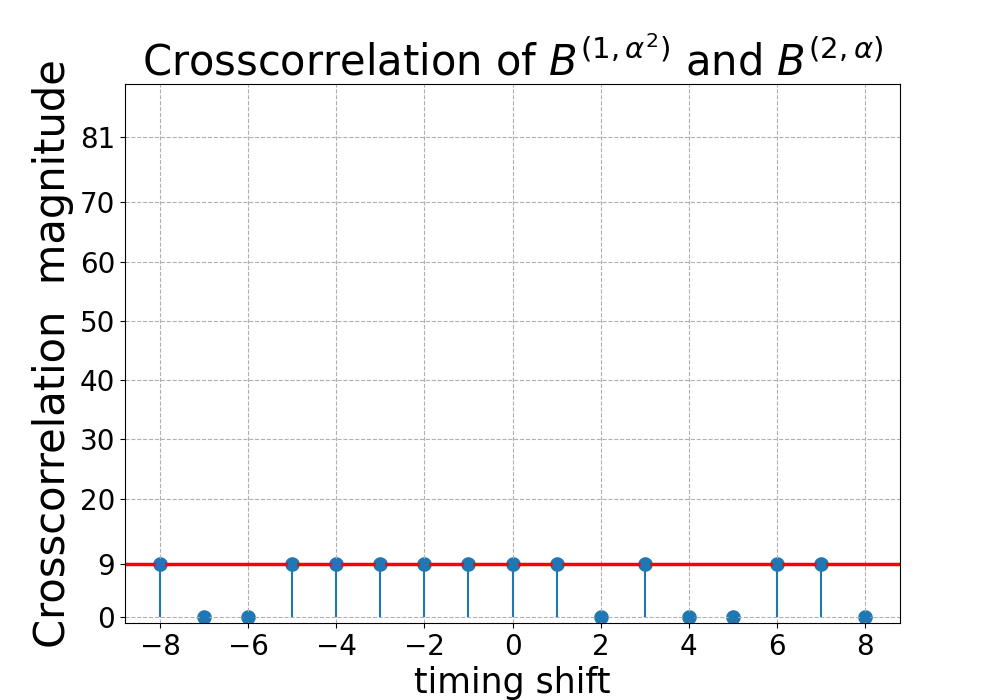}}
    \caption{The cross-correlation magnitude distributions of $\mathbf{B}^{1,\alpha}$,$\mathbf{B}^{1,\alpha^2}$ and $\mathbf{B}^{2,\alpha}$.} 
    \label{fig6} 
\end{figure}
\end{example}

\subsection{The third construction of aperiodic QCSSs}
Let $a,b\in \Bbb Z$. Construct a set 
\begin{eqnarray}\label{eqn-5}
\mathcal{D}=\{\mathbf{D}^{a,b}: 0\leq a\leq q-2, 0\leq b\leq q+1\}
\end{eqnarray}
containing $M=(q-1)(q+2)$ complementary sequences, where each matrix 
$\mathbf{D}^{a,b}=\left[\mathbf{s}_{0}^{a,b},\mathbf{s}_{1}^{a,b},\cdots,\mathbf{s}_{q-1}^{a,b}\right]^{T}$
consists of $K=q$ sequences of length $N=q+1$  defined by 
\begin{eqnarray*}
&\mathbf{s}_{k}^{a,b}=(\mathbf{s}_{k}^{a,b}(t))_{t=0}^{q}, \\~~ &\mathbf{s}_{k}^{a,b}(t)=\chi_1\left(d_k\tr_{q^2/q}(\beta^{a(q+1)+t})\right)\zeta_{q+2}^{bt}, ~ 0\leq k\leq q-1.
\end{eqnarray*}
 Clearly, the alphabet size of $\mathcal{D}$ is $p(q+2)$.
\begin{theorem}\label{444}
Let $q=p^n$ for a prime $p$ and a positive integer $n$. Then the set $\mathcal{D}$ defined in (\ref{eqn-5}) is an aperiodic $((q-1)(q+2), q, q+1, q)$-QCSS which is asymptotically optimal with the respect to the bound in (\ref{zl2}).
\end{theorem}

\begin{IEEEproof}Let $0\leq a_1,a_2\leq q-2, 0\leq b_1,b_2\leq q+1$.
For any two complementary sequences $\mathbf{D}^{a_{1},b_{1}}$, $\mathbf{D}^{a_{2},b_{2}}$ in $\mathcal{D}$ and $\tau\in[0,q]$, we have
\begin{eqnarray*}
\nonumber &
&T_{\mathbf{D}^{a_{1},b_{1}},\mathbf{D}^{a_{2},b_{2}}}(\tau)\\
\nonumber 
&=&\sum_{k=0}^{q-1}T_{\mathbf{s}_{k}^{a_{1},b_{1}},\mathbf{s}_{k}^{a_{2},b_{2}}}(\tau)
\\
&=&\sum_{k=0}^{q-1}\sum_{t=0}^{q-\tau}\chi_1\left(d_k\tr_{q^2/q}(\beta^{a_1(q+1)+t})\right)\zeta_{q+2}^{b_1t}
\\&&\overline{\chi}_1\left(d_k\tr_{q^2/q}(\beta^{a_2(q+1)+t+\tau})\right)\zeta_{q+2}^{-b_2(t+\tau)}\\
&=&\zeta_{q+2}^{-b_2\tau}\sum_{t=0}^{q-\tau}\zeta_{q+2}^{(b_1-b_2)t}\sum_{k=0}^{q-1}\chi_1\Big(d_k\big(\tr_{q^2/q}(\beta^{a_1(q+1)+t})
\\&&-\tr_{q^2/q}(\beta^{a_2(q+1)+t+\tau})\big)\Big).
\end{eqnarray*}
We divide into the following cases to determine the value distribution of $T_{\mathbf{D}^{a_{1},b_{1}},\mathbf{D}^{a_{2},b_{2}}}(\tau)$.

{Case 1}: If $\tau=0$, $a_1=a_2$ and $b_1\neq b_2$, then  
\begin{eqnarray*}
 &
&T_{\mathbf{D}^{a_{1},b_{1}},\mathbf{D}^{a_{2},b_{2}}}(\tau)\\
&=&\sum_{k=0}^{q-1}\sum_{t=0}^{q}\zeta_{q+2}^{(b_1-b_2)t}
=q\left(\sum_{t=0}^{q+1}\zeta_{q+2}^{(b_1-b_2)t}-\zeta_{q+2}^{(b_1-b_2)(q+1)}\right)\\
&=&-q\zeta_{q+2}^{(b_1-b_2)(q+1)},
\end{eqnarray*}
where $1\leq \vert b_1-b_2 \vert \leq q+1$ and 
$$\sum_{t=0}^{q+1}\zeta_{q+2}^{(b_1-b_2)t}=\frac{\zeta_{q+2}^{(b_1-b_2)0}-\zeta_{q+2}^{(b_1-b_2)(q+2)}}{1-\zeta_{q+2}^{b_1-b_2}}=0.$$

{Case 2}: If $\tau\neq0$ and $a_1=a_2$, then
\begin{eqnarray*}
 \begin{split}
&T_{\mathbf{D}^{a_{1},b_{1}},\mathbf{D}^{a_{2},b_{2}}}(\tau)=\zeta_{q+2}^{-b_2\tau}\sum_{t=0}^{q-\tau}\zeta_{q+2}^{(b_1-b_2)t}\times\\
&\sum_{k=0}^{q-1}\chi_1\Big(d_k\big(\tr_{q^2/q}(\beta^{a_1(q+1)+t})-\tr_{q^2/q}(\beta^{a_1(q+1)+t+\tau})\big)\Big)\\
&=\zeta_{q+2}^{-b_2\tau}\sum_{t=0}^{q-\tau}\zeta_{q+2}^{(b_1-b_2)t} \times\\
&\sum_{k=0}^{q-1}\chi_1\Big(d_k\big(\tr_{q^2/q}(\beta^{a_1(q+1)+t}(1-\beta^{\tau}))\big)\Big).
\end{split}
\end{eqnarray*}
Since $1\leq \tau \leq q$, we know that the equation $\tr_{q^2/q}\left(\beta^{a_1(q+1)+t}(1-\beta^{\tau})\right)=0$ with variable $t$ has at most one solution for $0\leq t \leq q-\tau$ by Lemma \ref{lem-2}.

{Subcase 2.1}: If $\tr_{q^2/q}\left(\beta^{a_1(q+1)+t}(1-\beta^{\tau})\right)=0$  has a unique solution $0\leq t_0 \leq q-\tau$, then 
\begin{eqnarray*}
\nonumber &
&T_{\mathbf{D}^{a_{1},b_{1}},\mathbf{D}^{a_{2},b_{2}}}(\tau)\\
&=&\zeta_{q+2}^{-b_2\tau}q\zeta_{q+2}^{(b_1-b_2){t_0}}+\zeta_{q+2}^{-b_2\tau}\sum_{\substack{t=0\\ t \neq t_0}}^{q-\tau}\zeta_{q+2}^{(b_1-b_2)t}
\\&&\sum_{k=0}^{q-1}\chi_1\left(d_k\tr_{q^2/q}\left(\beta^{a_1(q+1)+t}(1-\beta^{\tau})\right)\right)\\
&=&\zeta_{q+2}^{-b_2\tau}q\zeta_{q+2}^{(b_1-b_2){t_0}}+\zeta_{q+2}^{-b_2\tau}\sum_{\substack{t=0\\ t \neq t_0}}^{q-\tau}\zeta_{q+2}^{(b_1-b_2)t}\\
&&\sum_{y\in\gf_q}\chi_1\left(y\tr_{q^2/q}\left(\beta^{a_1(q+1)+t}(1-\beta^{\tau})\right)\right)\\
&=&q\zeta_{q+2}^{(b_1-b_2){t_0}-b_2\tau},
\end{eqnarray*}
where the third equation holds owing to the orthogonality relation of additive characters.

{Subcase 2.2}: If $\tr_{q^2/q}\left(\beta^{a_1(q+1)+t}(1-\beta^{\tau})\right)=0$  has no solution for $0\leq t \leq q-\tau$, then we have
 \begin{eqnarray*} 
 \begin{split}
&T_{\mathbf{D}^{a_{1},b_{1}},\mathbf{D}^{a_{2},b_{2}}}(\tau)\\
&=\zeta_{q+2}^{-b_2\tau}\sum_{t=0}^{q-\tau}\zeta_{q+2}^{(b_1-b_2)t}\sum_{y\in\gf_q}\chi_{1}\left(y\tr_{q^2/q}(\beta^{a_1(q+1)+t}(1-\beta^{\tau}))\right)\\
&=0,
\end{split}
\end{eqnarray*}
by the orthogonality relation of additive characters.

{Case 3}: If $\tau\in[0,q]$ and $a_1\neq a_2$, then
\begin{eqnarray*}
\begin{split}
&T_{\mathbf{D}^{a_{1},b_{1}},\mathbf{D}^{a_{2},b_{2}}}(\tau)\\
&=\zeta_{q+2}^{-b_2\tau}\sum_{t=0}^{q-\tau}\zeta_{q+2}^{(b_1-b_2)t}\\
&\sum_{k=0}^{q-1}\chi_1\left(d_k\left(\tr_{q^2/q}\left(\beta^{a_1(q+1)+t}(1-\beta^{(a_2-a_1)(q+1)+\tau})\right)\right)\right).
\end{split}
\end{eqnarray*}
Since $0\leq \tau \leq q$ and $0\leq a_1, a_2\leq q-2$ and $a_1\neq a_2$, it is easy to deduce $1-\beta^{(a_2-a_1)(q+1)+\tau} \neq 0$.
 Thus the equation $\tr_{q^2/q}\left(\beta^{a_1(q+1)+t}(1-\beta^{(a_1-a_2)(q+1)+\tau})\right)=0 $ with variable $t$ has at most one solution for $0\leq t \leq q-\tau$ by Lemma \ref{222}. 
Similarly to {Case 2} above, we have $T_{\mathbf{D}^{a_{1},b_{1}},\mathbf{D}^{a_{2},b_{2}}}(\tau)=q\zeta_{q+2}^{(b_1-b_2){t_0}-b_2\tau}$ if $\tr_{q^2/q}\left(\beta^{a_1(q+1)+t}(1-\beta^{(a_2-a_1)(q+1)+\tau})\right)=0$ has  a unique solution $t_0\in [0,q-\tau]$ and $T_{\mathbf{D}^{a_{1},b_{1}},\mathbf{D}^{a_{2},b_{2}}}(\tau)=0$ if $\tr_{q^2/q}\left(\beta^{a_1(q+1)+t}(1-\beta^{(a_2-a_1)(q+1)+\tau})\right)=0$ has no solution for $0\leq t \leq q-\tau$.

Summarizing the above three cases, we have
\begin{eqnarray*}
\vert T_{\mathbf{D}^{a_{1},b_{1}},\mathbf{D}^{a_{2},b_{2}}}(\tau)\vert\in\{0,q\}
\end{eqnarray*}
and the maximum aperiodic correlation magnitude $\theta_{\max}$ of $\mathcal{D}$ is $q$.

Finally, we show that the aperiodic QCSS $\mathcal{D}$  is asymptotically optimal according to
the bound in (\ref{zl2}). Since $\mathcal{D}$ is an aperiodic $((q-1)(q+2), q, q+1, q)$-QCSS, according to the lower bound in (\ref{zl2}), we have 
\begin{eqnarray*}
\theta_{\textup{opt}}=\sqrt{q(q+1)\left(1-2\sqrt{\frac{q}{3(q-1)(q+2)}}\right)}.
\end{eqnarray*}
It is easy to see that 
\begin{eqnarray*}
\lim_{q\rightarrow+\infty}\rho_1
=\lim_{q\rightarrow+\infty}\frac{q}{\sqrt{q(q+1)\left(1-2\sqrt{\frac{q}{3(q-1)(q+2)}}\right)}}=1,
\end{eqnarray*}
which completes the proof of this theorem.
\end{IEEEproof}

\begin{example}\label{example5}
Let $p=3$ and $n=2$. 
Then the QCSS $\mathcal{D}$ in Theorem \ref{444} is an aperiodic $(88, 9, 10, 9)$-QCSS with alphabet $\{e^{2\pi\sqrt{-1}i/33}: i\in[0,32]\}$, which is verified by Magma program. By Python program, we respectively show the auto-correlation and cross-correlation magnitude distributions of $\mathbf{D}^{1,1}$, $\mathbf{D}^{1,2}$ and $\mathbf{D}^{2,1}$ in Fig. \ref{fig9} and Fig. \ref{fig10}, where the red line stands for the maximum correlation magnitude. 

\begin{figure}[H]
    \centering
		\includegraphics[width=0.3\linewidth]{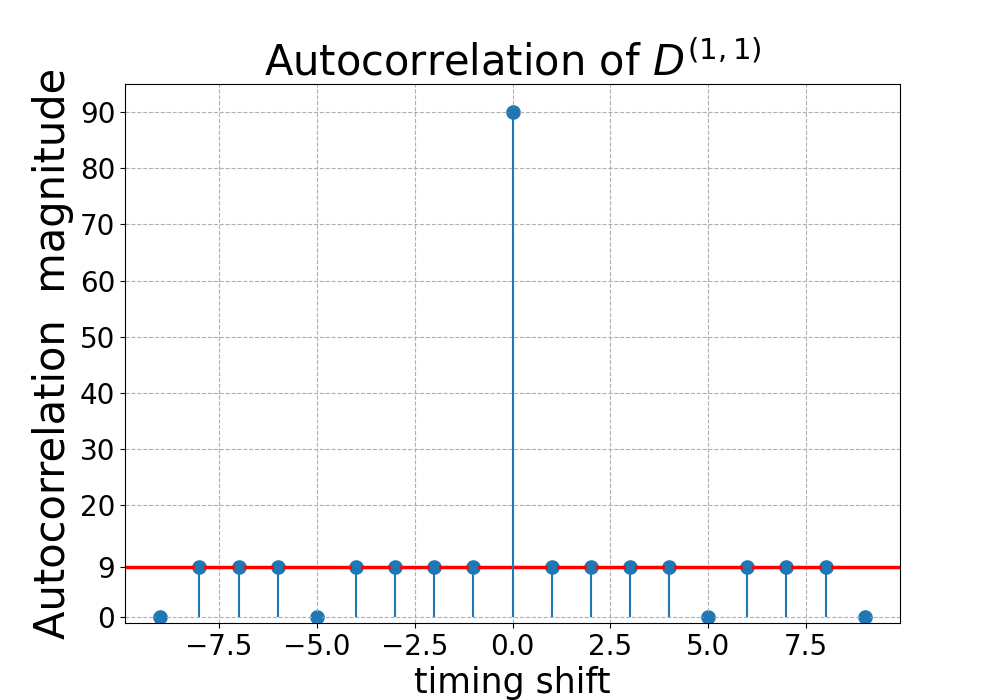}
    \hfill
		\includegraphics[width=0.3\linewidth]{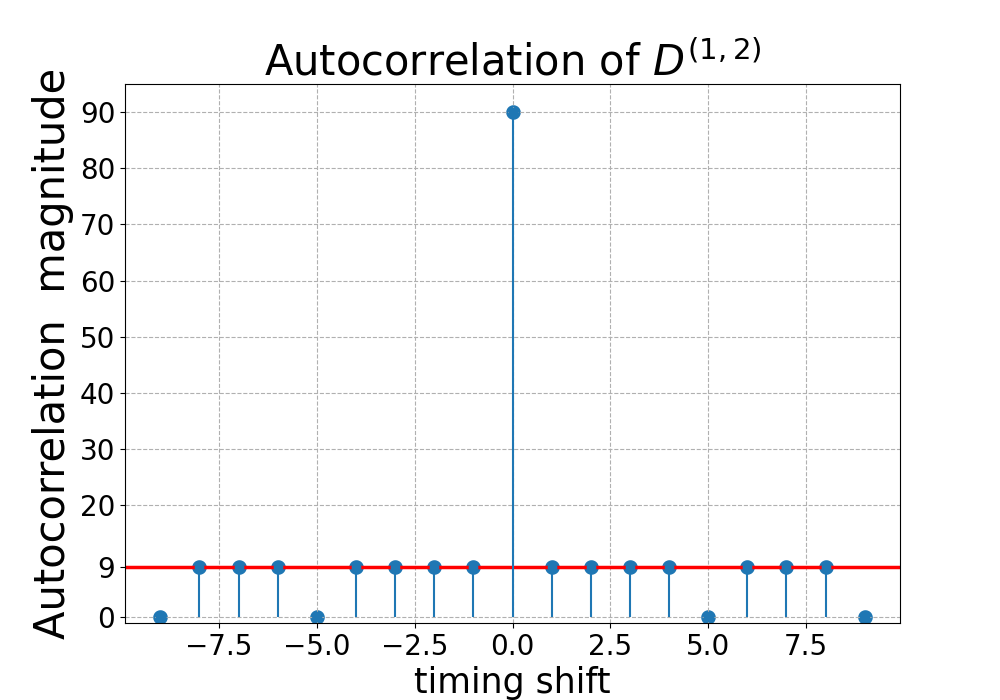}
    \hfill
		\includegraphics[width=0.3\linewidth]{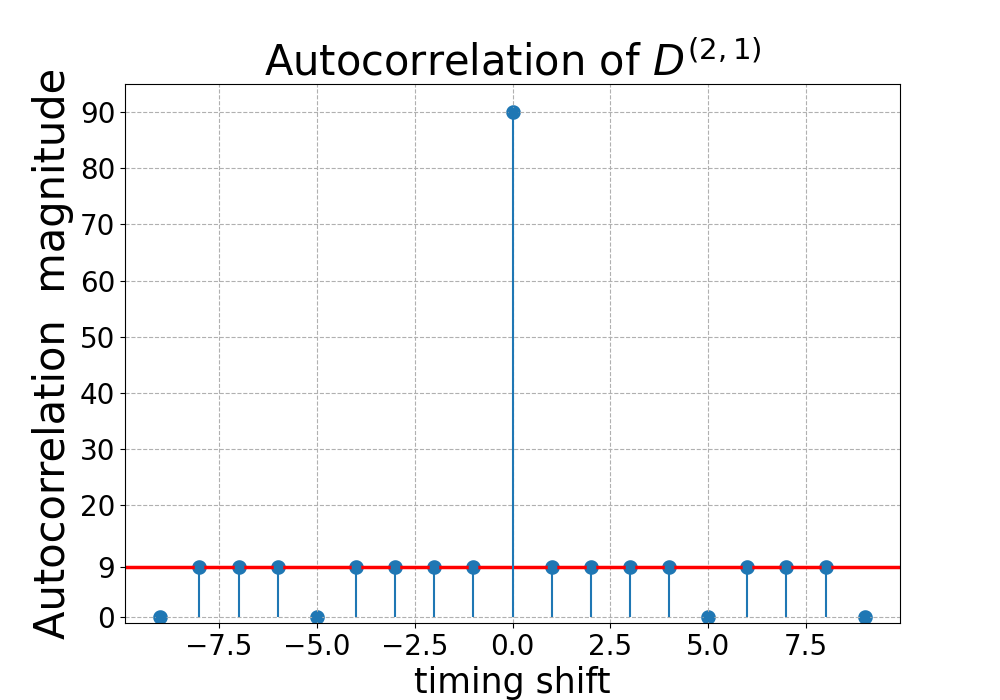}
    \caption{The auto-correlation magnitude distributions of $\mathbf{D}^{1,1}$,$\mathbf{D}^{1,2}$ and $\mathbf{D}^{2,1}$.} 
    \label{fig9} 
\end{figure}

\begin{figure}[H]
    \centering
		\includegraphics[width=0.3\linewidth]{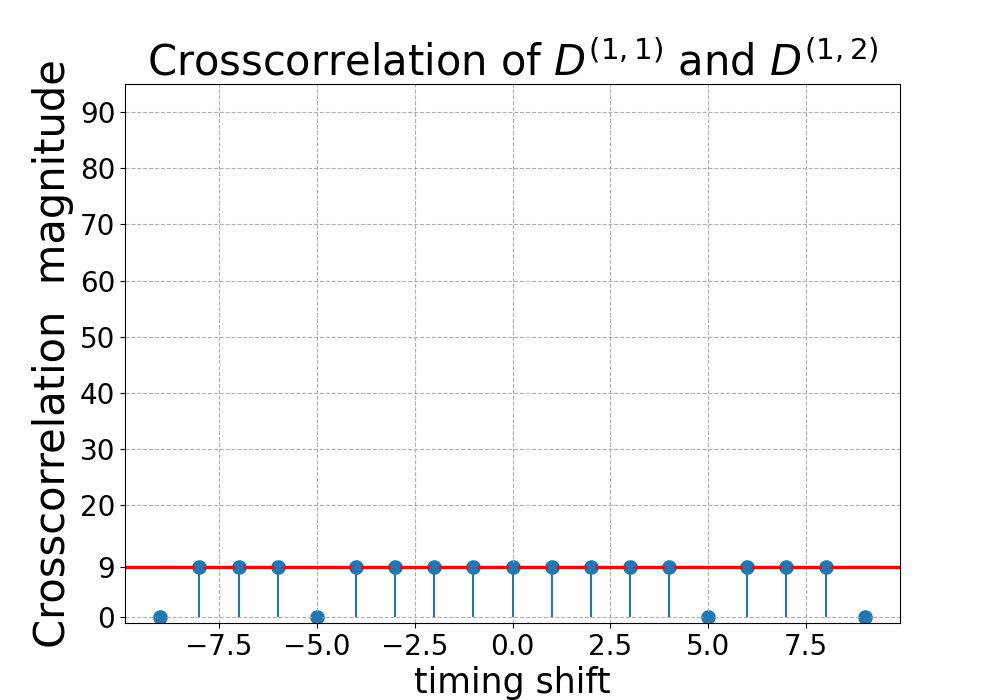}
    \hfill
		\includegraphics[width=0.3\linewidth]{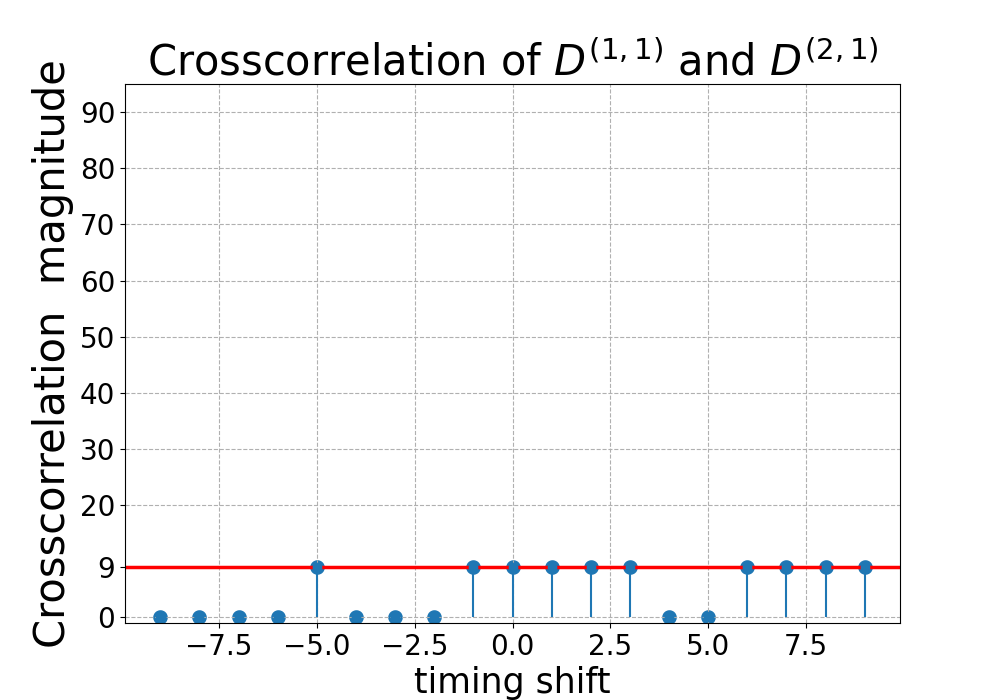}
    \hfill
		\includegraphics[width=0.3\linewidth]{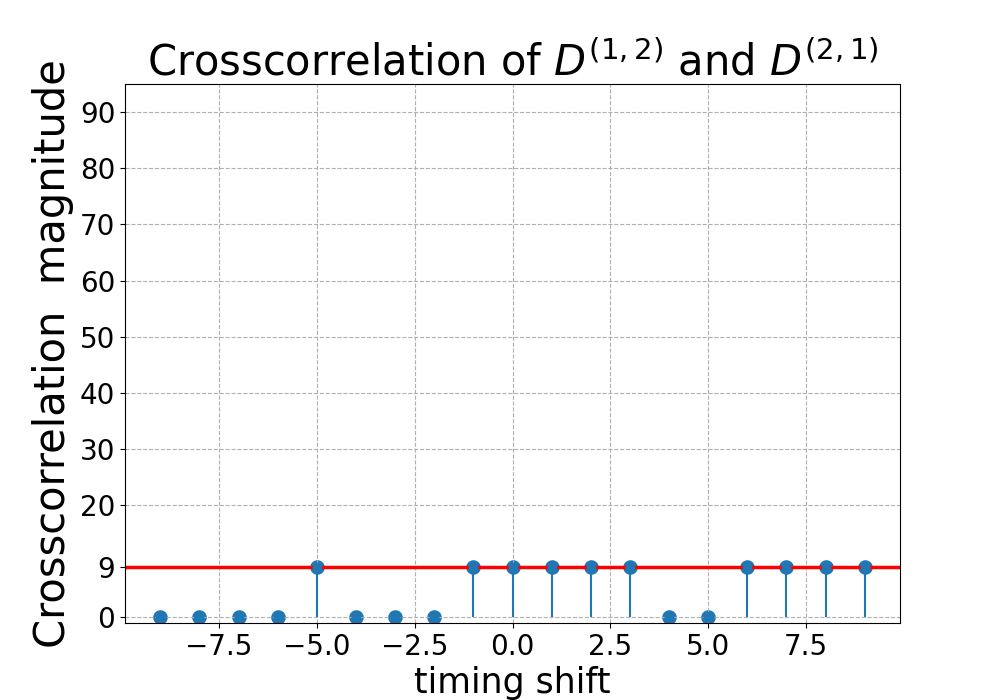}
    \caption{The cross-correlation magnitude distributions of $\mathbf{D}^{1,1}$,$\mathbf{D}^{1,2}$ and $\mathbf{D}^{2,1}$.} 
    \label{fig10} 
\end{figure} 
 
\end{example}

\subsection{The fourth construction of aperiodic QCSSs}
Let $0\leq e \leq q$ be an integer satisfying $\tr_{q^2/q}(\beta^e)=0$, where $\gf_{q^2}^*=\langle\beta \rangle$. Let $\phi(\cdot)$ be an arbitrary one-to-one mapping from $\gf_q$ to the set $\{0,1,\cdots,q-1\}$ such that $\phi(0)=0$. For instance, for $\gf_q^*=\langle\alpha\rangle$, we can define one of such mappings by $\phi(0)=0$ and $\phi(\alpha^i)=i+1$, where 
$0\leq i \leq q-2$.
Let $a,b\in \Bbb Z$. 
Construct a set 
\begin{eqnarray}\label{eqn-8}
\mathcal{E}=\{\mathbf{E}^{a,b}: 0\leq a\leq q-2, 0\leq b\leq q-2\}
\end{eqnarray}
containing $M=(q-1)^2$ complementary sequences, where each matrix
$\mathbf{E}^{a,b}=\left[\mathbf{s}_{0}^{a,b},\mathbf{s}_{1}^{a,b},\cdots,\mathbf{s}_{q-2}^{a,b}\right]^{T}$
consists of $K=q-1$ sequences of length $N=q-1$ defined by 
\begin{eqnarray*}
&\mathbf{s}_{k}^{a,b}=(\mathbf{s}_{k}^{a,b}(t))_{t=0}^{q-2},\\ ~~ &\mathbf{s}_{k}^{a,b}(t)=\zeta_{q-1}^{k\phi\left(\tr_{q^2/q}(\beta^{e+a(q+1)+t+1})\right)+bt}, ~ 0\leq k\leq q-2.
\end{eqnarray*}
 Thus the alphabet size of $\mathcal{E}$ is $q-1$.
\begin{theorem}\label{777}
Let $q=p^n>2$ for a prime $p$ and a positive integer $n$. Then the set $\mathcal{E}$ defined in (\ref{eqn-8}) is an aperiodic $((q-1)^2, q-1, q-1, q-1)$-QCSS with alphabet size $q-1$ which is asymptotically optimal with the respect to the bound in (\ref{zl2}) .
\end{theorem}

\begin{IEEEproof}
Let $a_1,a_2,b_1,b_2$ be integers such that $0\leq a_1,a_2\leq q-2, 0\leq b_1,b_2\leq q-2$.
For any two complementary sequences $\mathbf{E}^{a_{1},b_{1}}$, $\mathbf{E}^{a_{2},b_{2}}$ in $\mathcal{E}$ and $\tau\in[0,q-2]$, we have
\begin{eqnarray*} 
&
&T_{\mathbf{E}^{a_{1},b_{1}},\mathbf{E}^{a_{2},b_{2}}}(\tau) 
=\sum_{k=0}^{q-2}T_{\mathbf{s}_{k}^{a_{1},b_{1}},\mathbf{s}_{k}^{a_{2},b_{2}}}(\tau)
\\
&=&\sum_{k=0}^{q-2}\sum_{t=0}^{q-2-\tau}\zeta_{q-1}^{k\phi\left(\tr_{q^2/q}(\beta^{e+a_1(q+1)+t+1})\right)+b_1t}
\\&&\zeta_{q-1}^{-k\phi\left(\tr_{q^2/q}(\beta^{e+a_2(q+1)+t+\tau+1})\right)-b_2(t+\tau)}\\
&=&\zeta_{q-1}^{-b_2\tau}\sum_{t=0}^{q-2-\tau}\zeta_{q-1}^{(b_1-b_2)t}\sum_{k=0}^{q-2}
\zeta_{q-1}^{kg(t)},
\end{eqnarray*}
where $g(t):=\phi\left(\tr_{q^2/q}(\beta^{e+a_1(q+1)+t+1})\right)-\phi\left(\tr_{q^2/q}(\beta^{e+a_2(q+1)+t+\tau+1})\right)$.
Note that $$\tr_{q^2/q}(\beta^{e+a_1(q+1)+t+1})=\beta^{a_1(q+1)}\tr_{q^2/q}(\beta^{e+t+1}).$$
By Lemma \ref{lem-2}, $\tr_{q^2/q}(\beta^j)=0$ has at most one zero for $j\in \{e,e+1,\cdots,e+(q-1-\tau)\}$.
Hence $\tr_{q^2/q}(\beta^{e+a_1(q+1)+t+1})\neq0$ for  $0\leq t\leq q-2-\tau$ as  $\tr_{q^2/q}(\beta^e)=0$. 
Similarly, we also have $\tr_{q^2/q}(\beta^{e+a_2(q+1)+t+\tau+1})\neq 0$ as $t+\tau+2<q+1$ for  $0\leq t\leq q-2-\tau$.
Then $0 \leq |g(t)| \leq q-2$.
Let $h(t):=\tr_{q^2/q}(\beta^{e+a_1(q+1)+t+1})-\tr_{q^2/q}(\beta^{e+a_2(q+1)+t+\tau+1})$.
Since $\phi(\cdot)$ is an arbitrary one-to-one mapping, we have $g(t)=0$ if and only if $h(t)=0$.
We divide into the following cases to determine the value distribution of $T_{\mathbf{E}^{a_{1},b_{1}},\mathbf{E}^{a_{2},b_{2}}}(\tau)$.

{Case 1}: If $\tau=0$, $a_1=a_2$ and $b_1\neq b_2$, then  
$T_{\mathbf{E}^{a_{1},b_{1}},\mathbf{E}^{a_{2},b_{2}}}(\tau)=\sum_{k=0}^{q-2}\sum_{t=0}^{q-2}\zeta_{q-1}^{(b_1-b_2)t}=0$
as $1\leq \vert b_1-b_2 \vert \leq q-2$.

{Case 2}: If $\tau\neq0$ and $a_1= a_2$, then  
\begin{eqnarray*}
T_{\mathbf{E}^{a_{1},b_{1}},\mathbf{E}^{a_{2},b_{2}}}(\tau)=\zeta_{q-1}^{-b_2\tau}\sum_{t=0}^{q-2-\tau}\zeta_{q-1}^{(b_1-b_2)t}\sum_{k=0}^{q-2}\zeta_{q-1}^{kg(t)}.
\end{eqnarray*}
Note that $h(t)=\tr_{q^2/q}(\beta^{e+a_1(q+1)+t+1}(1-\beta^\tau))=0$ with $0\leq t \leq q-2-\tau$ has at most one solution $t_0$ satisfying $h(t_0)=0$ by Lemma \ref{lem-2}.

{Subcase 2.1}: If $h(t)=0$ has a unique solution $0\leq t_0 \leq q-2-\tau$, then $g(t)$ also has the unique  solution $t_0$ and
\begin{eqnarray*}
\begin{split}
&T_{\mathbf{E}^{a_{1},b_{1}},\mathbf{E}^{a_{2},b_{2}}}(\tau)\\
&=\zeta_{q-1}^{-b_2\tau}(q-1)\zeta_{q-1}^{(b_1-b_2)_{t_0}}+\zeta_{q-1}^{-b_2\tau}\sum_{\substack{t=0\\ t \neq t_0}}^{q-2-\tau}\zeta_{q-1}^{(b_1-b_2)t}\sum_{k=0}^{q-2}\zeta_{q-1}^{kg(t)}\\
&=\zeta_{q-1}^{-b_2\tau}(q-1)\zeta_{q-1}^{(b_1-b_2)_{t_0}}
\end{split}
\end{eqnarray*}
as $0 < |g(t)| \leq q-2$ for $t\neq t_0$.

{Subcase 2.2}: If $h(t)=0$ has no solution for $0\leq t \leq q-2-\tau$, then $0 < |g(t)| \leq q-2$ and
\begin{eqnarray*}
T_{\mathbf{E}^{a_{1},b_{1}},\mathbf{E}^{a_{2},b_{2}}}(\tau)
=\zeta_{q-1}^{-b_2\tau}\sum_{t=0}^{q-2-\tau}\zeta_{q-1}^{(b_1-b_2)t}\sum_{k=0}^{q-2}\zeta_{q-1}^{kg(t)}
=0.
\end{eqnarray*}

{Case 3}: If $\tau\in[0,q-2]$ and $a_1\neq a_2$, then 
\begin{eqnarray*}
T_{\mathbf{E}^{a_{1},b_{1}},\mathbf{E}^{a_{2},b_{2}}}(\tau)=\zeta_{q-1}^{-b_2\tau}\sum_{t=0}^{q-2-\tau}\zeta_{q-1}^{(b_1-b_2)t}\sum_{k=0}^{q-2}\zeta_{q-1}^{kg(t)}.
\end{eqnarray*}
Note that $h(t)=\tr_{q^2/q}(\beta^{e+a_1(q+1)+t+1}(1-\beta^{(a_1-a_2)(q+1)+\tau}))$, where  $q^2-1\nmid\left((a_1-a_2)(q+1)+\tau\right)$ as $0<|a_1-a_2|\leq q-2$ and $\tau\in[0,q-2]$.
By Lemma \ref{lem-2}, $h(t)$ has at most one zero for $0\leq t\leq q-2-\tau$.
Then $g(t)$ also has at most one zero for $0\leq t\leq q-2-\tau$.

{Subcase 3.1}: If $h(t)=0$ has a unique solution $t_0$ for  $0\leq t_0\leq q-2-\tau$, then  $g(t)$ also has the unique  solution $t_0$ and
\begin{eqnarray*}
& T_{\mathbf{E}^{a_{1},b_{1}},\mathbf{E}^{a_{2},b_{2}}}(\tau)\\
&=\zeta_{q-1}^{-b_2\tau}(q-1)\zeta_{q-1}^{(b_1-b_2)_{t_0}}+\zeta_{q-1}^{-b_2\tau}\sum\limits_{\substack{t=0\\ t \neq t_0}}^{q-2-\tau}\zeta_{q-1}^{(b_1-b_2)t}\sum\limits_{k=0}^{q-2}\zeta_{q-1}^{kg(t)}\\
&=\zeta_{q-1}^{-b_2\tau}(q-1)\zeta_{q-1}^{(b_1-b_2)_{t_0}}
\end{eqnarray*}
as $0 < |g(t)| \leq q-2$ for $t\neq t_0$.

{Subcase 3.2}: If $h(t)=0$ has no solution for $0\leq t \leq q-2-\tau$, then $0 < |g(t)| \leq q-2$ and
\begin{eqnarray*}
T_{\mathbf{E}^{a_{1},b_{1}},\mathbf{E}^{a_{2},b_{2}}}(\tau)
=\zeta_{q-1}^{-b_2\tau}\sum_{t=0}^{q-2-\tau}\zeta_{q-1}^{(b_1-b_2)t}\sum_{k=0}^{q-2}\zeta_{q-1}^{kg(t)}=0.
\end{eqnarray*}

Summarizing the above three cases, we have
\begin{eqnarray*}
\vert T_{\mathbf{E}^{a_{1},b_{1}},\mathbf{E}^{a_{2},b_{2}}}(\tau)\vert\in\{0,q-1\}.
\end{eqnarray*}
Note that $T_{\mathbf{E}^{a_{1},b_{1}},\mathbf{E}^{a_{2},b_{2}}}(\tau)$ can not be zero for all the cases.
Otherwise, $\mathcal{E}$ is a PCSS whose set size is at most $q-1$, which contradicts with  $|\mathcal{E}|=(q-1)^2$.
Thus the maximum aperiodic correlation magnitude of $\mathcal{E}$ is $q-1$.
 Since $\mathcal{E}$ is an aperiodic $((q-1)^2, q-1, q-1, q-1)$-QCSS, according to the lower bound in (\ref{zl2}), we have 
\begin{eqnarray*}
\theta_{\textup{opt}}=\sqrt{(q-1)^2\left(1-2\sqrt{\frac{q-1}{3(q-1)^2}}\right)}.
\end{eqnarray*}
It is easy to see that 
\begin{eqnarray*}
\lim_{q\rightarrow+\infty}\rho_1
=\lim_{q\rightarrow+\infty}\frac{q-1}{\sqrt{(q-1)^2\left(1-2\sqrt{\frac{q-1}{3(q-1)^2}}\right)}}=1,
\end{eqnarray*}
which completes the proof.
\end{IEEEproof}

\begin{figure}[H]
    \centering
		{\includegraphics[width=0.3\linewidth]{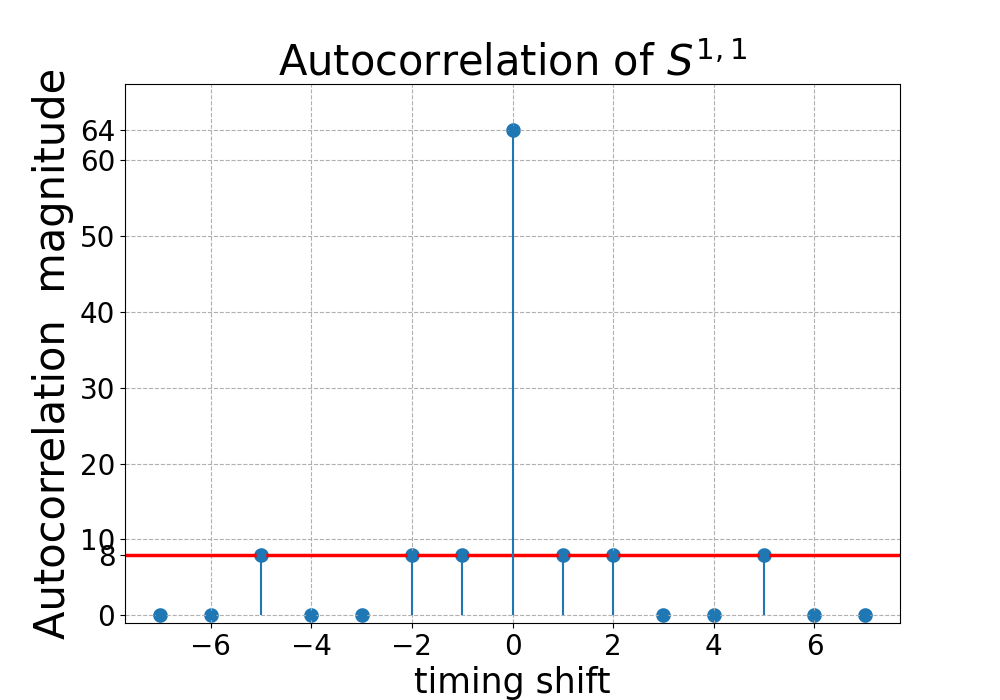}}
    \hfill
		{\includegraphics[width=0.3\linewidth]{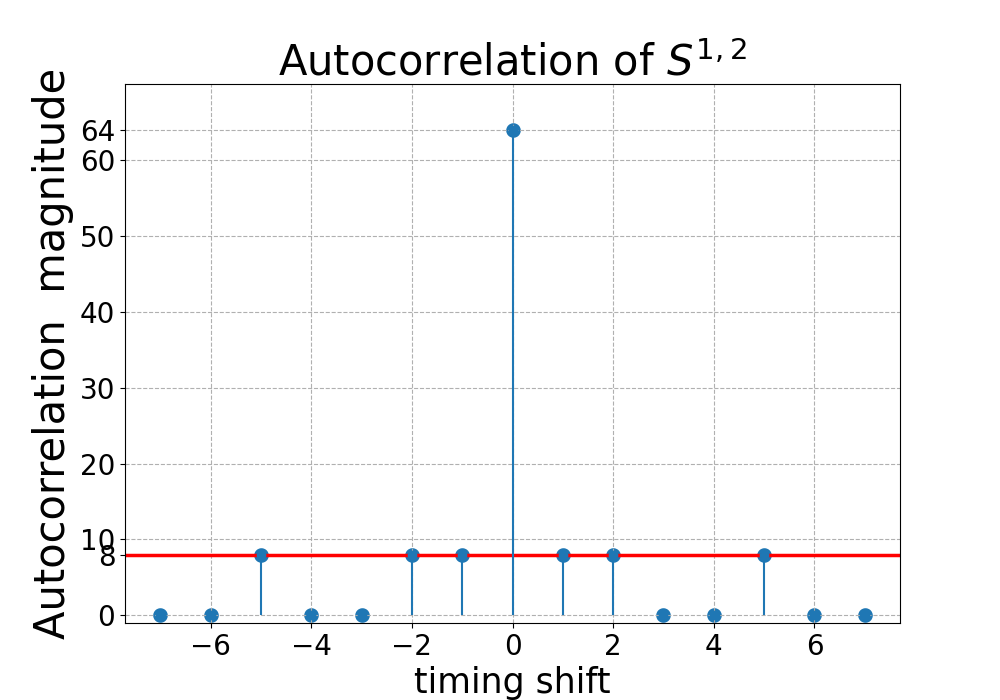}}
    \hfill
		{\includegraphics[width=0.3\linewidth]{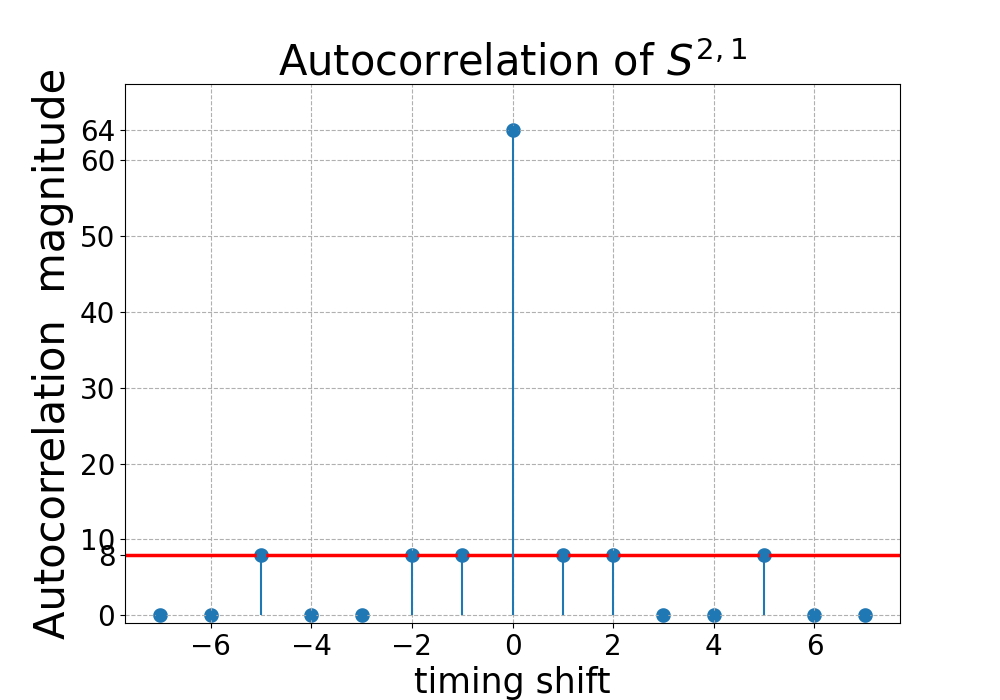}}
    \caption{The auto-correlation magnitude distributions of $\mathbf{E}^{1,1}$,$\mathbf{E}^{1,2}$ and $\mathbf{E}^{2,1}$.} 
    \label{fig11} 
\end{figure}

\begin{figure}[H]
    \centering
		{\includegraphics[width=0.3\linewidth]{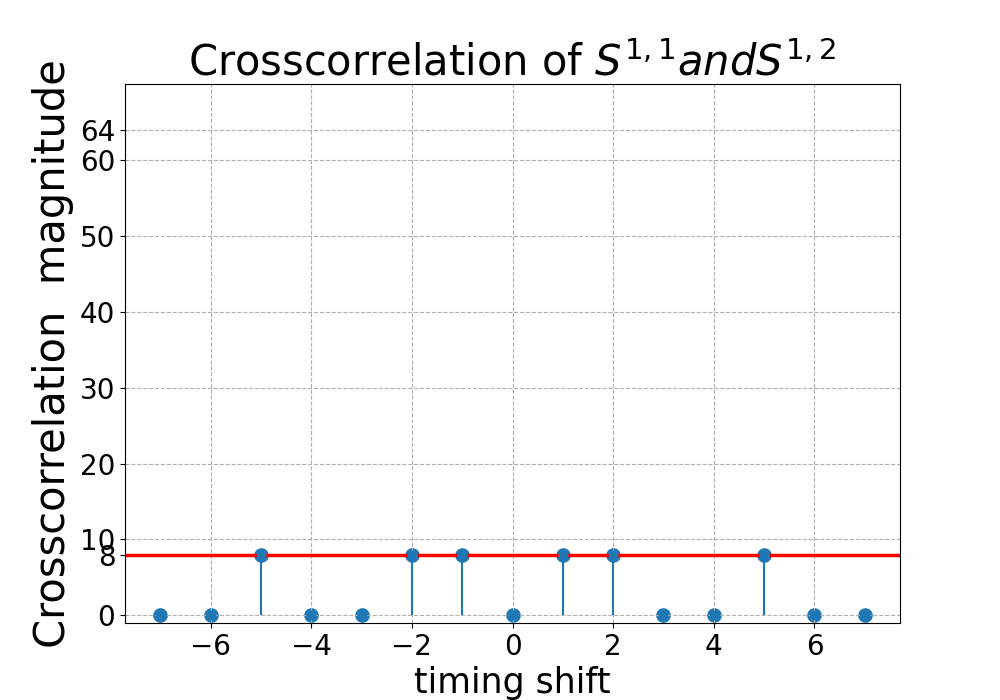}}
    \hfill
		{\includegraphics[width=0.3\linewidth]{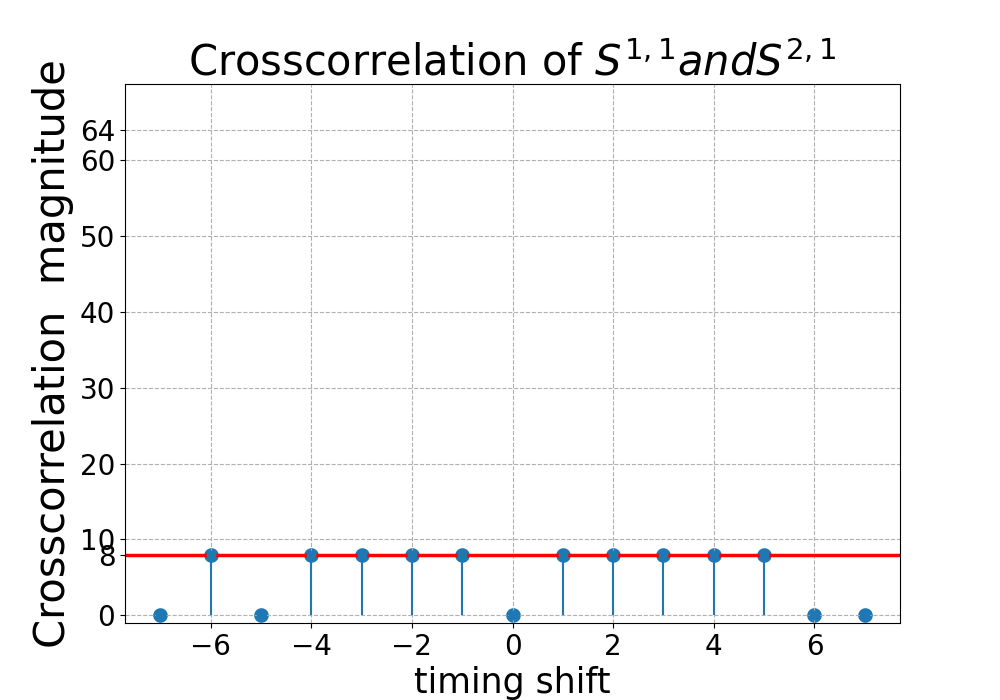}}
    \hfill
		{\includegraphics[width=0.3\linewidth]{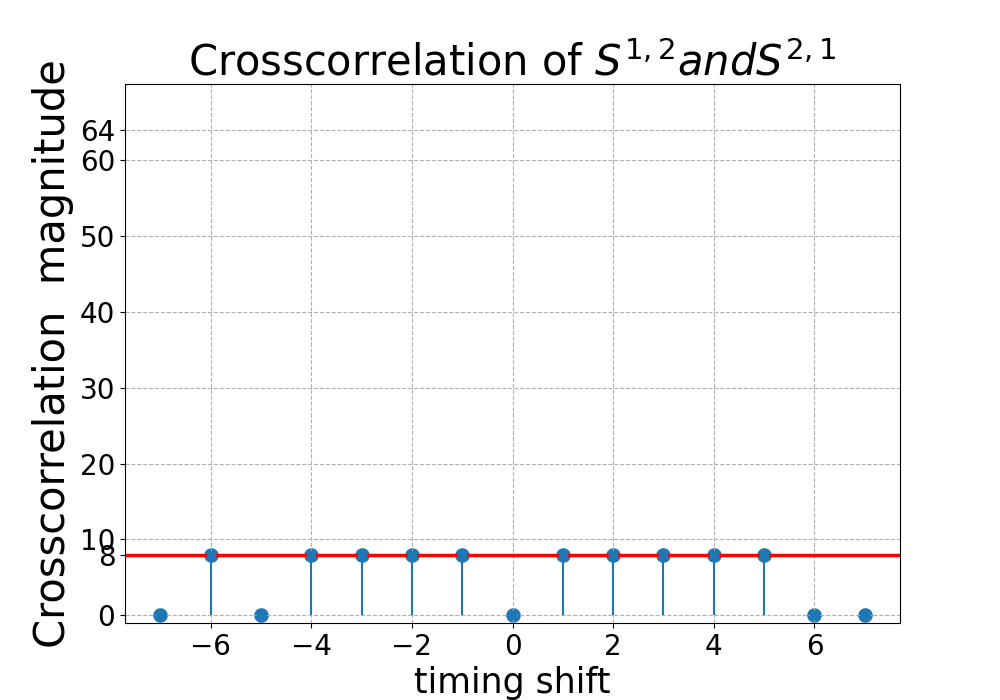}}
    \caption{The cross-correlation magnitude distributions of $\mathbf{E}^{1,1}$,$\mathbf{E}^{1,2}$ and $\mathbf{E}^{2,1}$.} 
    \label{fig12} 
\end{figure}

\begin{example}\label{example3}
Let $p=3$ and $n=2$. Define the one-to-one mapping $\phi(\cdot)$  by $\phi(0)=0$ and $\phi(\alpha^i)=i+1$, where 
$0\leq i \leq 7$.
Then the set $\mathcal{E}$ constructed in Theorem \ref{777} is an aperiodic $(64, 8, 8, 8)$-QCSS with alphabet $\{e^{2\pi\sqrt{-1}i/8}: i\in[0,7]\}$, which is verified by Magma program. By Python program, we respectively show the auto-correlation and cross-correlation magnitude distributions of $\mathbf{E}^{1,1}$, $\mathbf{E}^{1,2}$ and $\mathbf{E}^{2,1}$in Fig. \ref{fig11} and Fig. \ref{fig12}, where the red line stands for the maximum correlation magnitude. 
\end{example}

\subsection{The fifth construction of aperiodic QCSSs}
Let $\chi_{1}$ and $\varphi_{i}$ respectively be the canonical additive character and multiplicative character of $\gf_q$ defined in Section II. Construct a set 
\begin{eqnarray}\label{eqn-9}
\mathcal{F}=\{\mathbf{F}^{i,b}: 0\leq i\leq q-2, b\in \gf_q\}
\end{eqnarray}
 containing $M=q(q-1)$ complementary sequences, where each matrix
$\mathbf{F}^{i,b}=\left[\mathbf{f}_{0}^{i,b},\mathbf{f}_{1}^{i,b},\cdots,\mathbf{f}_{q-1}^{i,b}\right]^{T}$
consists of $K=q$ sequences of length $N=q-2$ defined as
\begin{eqnarray*}
&\mathbf{f}_{l}^{i,b}=(\mathbf{f}_{l}^{i,b}(t))_{t=0}^{q-3},\\ ~~ &\mathbf{f}_{l}^{i,b}(t)=\varphi_{i}(\alpha^t)\chi_{1}(d_l\alpha^{t}+bd_l), ~ 0\leq l\leq q-1.
\end{eqnarray*}
Then $\mathcal{F}$ has alphabet size $p(q-1)$.

\begin{theorem}\label{888}
Let $q=p^n$ for a  prime $p$ and a positive integer $n$ such that $q>2$. The the set $\mathcal{F}$ defined in (\ref{eqn-9}) is an aperiodic $(q(q-1), q, q-2, q)$-QCSS which is asymptotically optimal with the respect to the bound in  (\ref{zl2}) .
\end{theorem}
\begin{IEEEproof}
For any two  complementary sequences $\mathbf{F}^{i_{1},b_{1}}$, $\mathbf{F}^{i_{2},b_{2}}$ in $\mathcal{F}$ and $\tau\in[0,q-3]$, we have
\begin{eqnarray*}
 &
&T_{\mathbf{F}^{i_{1},b_{1}},\mathbf{F}^{i_{2},b_{2}}}(\tau)
=\sum_{l=0}^{q-1}T_{\mathbf{f}_{l}^{i_{1},b_{1}},\mathbf{f}_{l}^{i_{2},b_{2}}}(\tau)
\\
&=&\sum_{l=0}^{q-1}\sum_{t=0}^{q-3-\tau}\varphi_{i_1}(\alpha^t)\chi_1(d_l\alpha^t+b_1d_l)\overline{\varphi}_{i_2}(\alpha^{t+\tau})
\\&&\overline{\chi_1}(d_l\alpha^{t+\tau}+b_2d_l)
\\
&=&\sum_{l=0}^{q-1}\sum_{t=0}^{q-3-\tau}\overline{\varphi}_{i_2}(\alpha^\tau)\varphi_{i_1-i_2}(\alpha^t)\chi_1(d_l\alpha^t(1-\alpha^\tau)+
\\&&(b_1-b_2)d_l),
\end{eqnarray*}
where $0\leq i_1, i_2 \leq q-2$, $b_1, b_2\in \gf_q$.
We divide into the following cases to determine the value distribution of $T_{\mathbf{F}^{i_{1},b_{1}},\mathbf{F}^{i_{2},b_{2}}}(\tau)$.

{Case 1}: If $\tau=0$, then 
\begin{eqnarray*}
\begin{split}
&T_{\mathbf{F}^{i_{1},b_{1}},\mathbf{F}^{i_{2},b_{2}}}(\tau)
=\sum_{l=0}^{q-1}\sum_{t=0}^{q-3}\varphi_{i_1-i_2}(\alpha^t)\chi_{1}((b_1-b_2)d_l) \\
&=\sum_{y\in\gf_q}\sum_{t=0}^{q-3}\varphi_{i_1-i_2}(\alpha^t)\chi_{1}((b_1-b_2)y).
\end{split}
\end{eqnarray*}

{Subcase 1.1}: If $i_1=i_2$ and $b_1\neq b_2$, then 
\begin{eqnarray*}
T_{\mathbf{F}^{i_{1},b_{1}},\mathbf{F}^{i_{2},b_{2}}}(\tau)=\sum_{t=0}^{q-3}\sum_{y\in\gf_q}\chi_{1}((b_1-b_2)y)=0
\end{eqnarray*}by the orthogonality relation of additive characters.

{Subcase 1.2}: If $i_1\neq i_2$ and $b_1=b_2$, then 
\begin{eqnarray*}
T_{\mathbf{F}^{i_{1},b_{1}},\mathbf{F}^{i_{2},b_{2}}}(\tau)=\sum_{y\in\gf_q}\sum_{t=0}^{q-3}\varphi_{i_1-i_2}(\alpha^t)=-q\varphi_{i_1-i_2}(\alpha^{q-2})
\end{eqnarray*}
by the orthogonality relation of multiplicative characters.

{Subcase 1.3}: If $i_1\neq i_2$ and $b_1\neq b_2$, then 
\begin{eqnarray*}
T_{\mathbf{F}^{i_{1},b_{1}},\mathbf{F}^{i_{2},b_{2}}}(\tau)=\sum_{t=0}^{q-3}\varphi_{i_1-i_2}(\alpha^t)\sum_{y\in\gf_q}\chi_{1}((b_1-b_2)y)=0
\end{eqnarray*}
by the orthogonality relation of additive characters.

{Case 2}: If $\tau\neq0$, then 
\begin{eqnarray*}
\begin{split}
&T_{\mathbf{F}^{i_{1},b_{1}},\mathbf{F}^{i_{2},b_{2}}}(\tau)=\\
&
\sum_{t=0}^{q-3-\tau}\overline{\varphi}_{i_2}(\alpha^\tau)\varphi_{i_1-i_2}(\alpha^t)\sum_{y\in\gf_q}
\chi_1(y(\alpha^t(1-\alpha^\tau)+b_1-b_2)).
\end{split}
\end{eqnarray*}
Note that $1-\alpha^{\tau}\neq0$ for $\tau\neq0$. Thus we consider the following subcases.

Subcase 2.1: If $b_1-b_2=0$, then 
\begin{eqnarray*}
\nonumber &
&T_{\mathbf{F}^{i_{1},b_{1}},\mathbf{F}^{i_{2},b_{2}}}(\tau)\\
&=&\sum_{t=0}^{q-3-\tau}\overline{\varphi}_{i_2}(\alpha^\tau)\varphi_{i_1-i_2}(\alpha^t)\sum_{y\in\gf_q}\chi_1(y\alpha^t(1-\alpha^\tau))=0
\end{eqnarray*}
by the orthogonal relation of the  additive characters.

Subcase 2.2: If $b_1-b_2\neq0$, then 
\begin{eqnarray*}
T_{\mathbf{F}^{i_{1},b_{1}},\mathbf{F}^{i_{2},b_{2}}}(\tau)=\sum_{t=0}^{q-3-\tau}\overline{\varphi}_{i_2}(\alpha^\tau)\varphi_{i_1-i_2}(\alpha^t)\sum_{y\in\gf_q}\chi_1(yg(t)),
\end{eqnarray*}
where $g(t):=\alpha^{t}(1-\alpha^{\tau})+b_1-b_2$ has at most one root for $0\leq t \leq q-3-\tau$. 
If $g(t)$ has a unique root $0\leq t_0 \leq q-3-\tau$,
by the orthogonal relation of the  additive characters, we have
\begin{eqnarray*} 
\begin{split}
&T_{\mathbf{F}^{i_{1},b_{1}},\mathbf{F}^{i_{2},b_{2}}}(\tau)=
q\overline{\varphi}_{i_2}(\alpha^\tau)\varphi_{i_1-i_2}(\alpha^{t_{0}})+\\
&\sum_{\substack{t=0 t \neq t_0}}^{q-3-\tau}\overline{\varphi}_{i_2}(\alpha^\tau)\varphi_{i_1-i_2}(\alpha^t)
\sum_{y\in\gf_q}\chi_1(yg(t))\\
&=q\overline{\varphi}_{i_2}(\alpha^\tau)\varphi_{i_1-i_2}(\alpha^{t_{0}}).
\end{split}
\end{eqnarray*}
If $g(t)$ has no root for $0\leq t \leq q-3-\tau$, then $T_{\mathbf{F}^{i_{1},b_{1}},\mathbf{F}^{i_{2},b_{2}}}(\tau)=0$
by the orthogonality relation of additive characters.

Summarizing the above two cases, we have
\begin{eqnarray*}
\vert T_{\mathbf{F}^{i_{1},b_{1}},\mathbf{F}^{i_{2},b_{2}}}(\tau)\vert\in\{0,q\}.
\end{eqnarray*}
 Since $\mathcal{F}$ is an aperiodic $(q(q-1), q, q-2, q)$-QCSS, according to the lower bound in (\ref{zl2}), we have 
\begin{eqnarray*}
\theta_{\textup{opt}}=\sqrt{q(q-2)\left(1-2\sqrt{\frac{q}{3q(q-1)}}\right)}.
\end{eqnarray*}
It is easy to see that 
\begin{eqnarray*}
\lim_{q\rightarrow+\infty}\rho_1
=\lim_{q\rightarrow+\infty}\frac{q}{\sqrt{q(q-2)\left(1-2\sqrt{\frac{q}{3q(q-1)}}\right)}}=1.
\end{eqnarray*}
The desired conclusion follows.
\end{IEEEproof}

\begin{example}\label{example1} 
Let $p=2$ and $n=3$. Then the set $\mathcal{F}$ constructed in Theorem \ref{888} is an aperiodic $(56, 8, 6, 8)$-QCSS, which is verified by Magma program. By Python program, we respectively show the auto-correlation and cross-correlation magnitude distributions of $\mathbf{F}^{0,\alpha}$, $\mathbf{F}^{0,\alpha^2}$ and $\mathbf{F}^{1,\alpha^2}$ in Fig. \ref{fig13} and Fig. \ref{fig14}, where the red line stands for the maximum correlation magnitude.

\begin{figure}[H]
   \centering
		\includegraphics[width=0.3\linewidth]{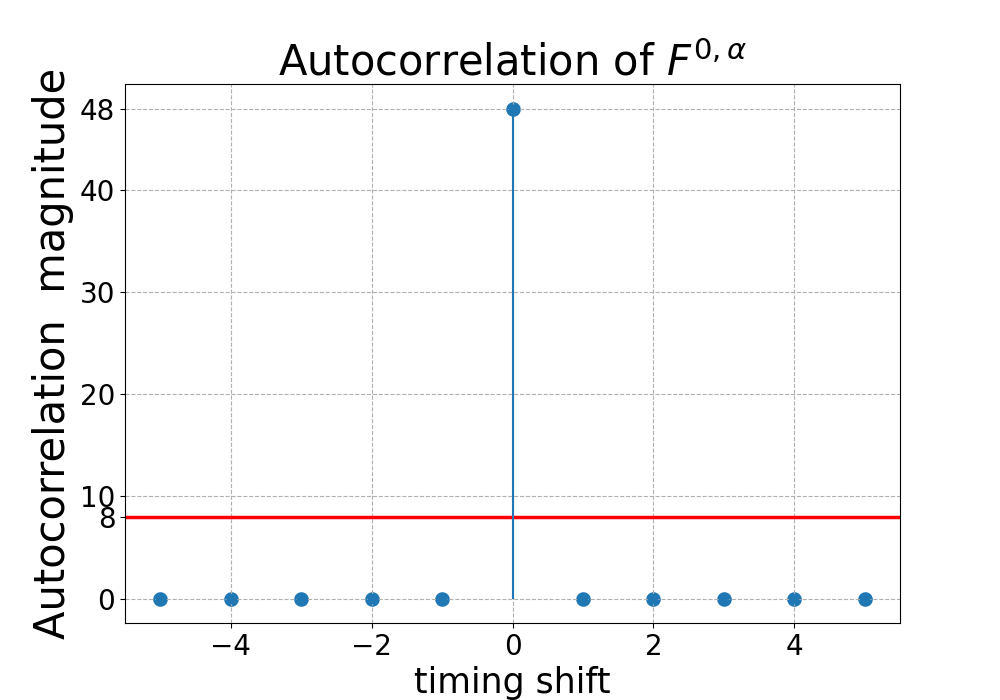}
    \hfill
		\includegraphics[width=0.3\linewidth]{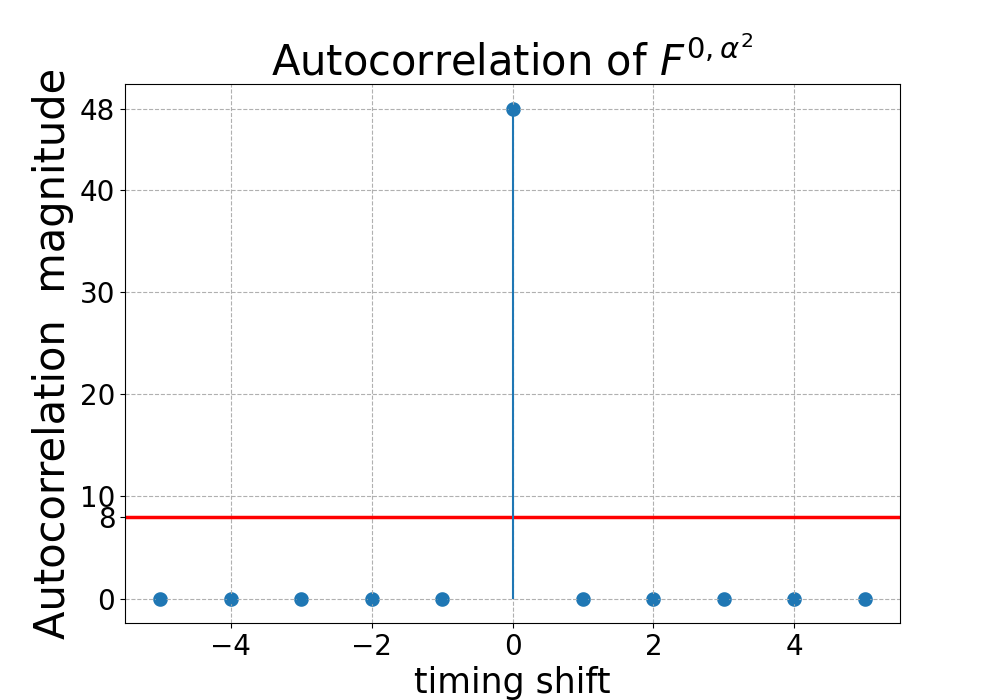}
    \hfill
		\includegraphics[width=0.3\linewidth]{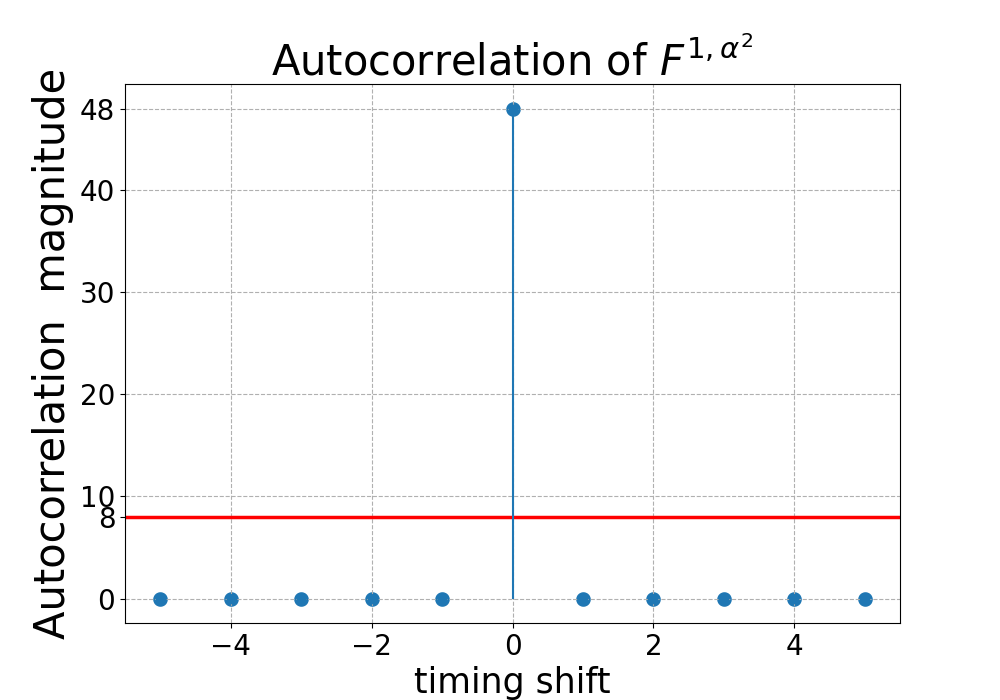}
    \caption{The auto-correlation magnitude distributions of $\mathbf{F}^{0,\alpha}$,$\mathbf{F}^{0,\alpha^2}$ and $\mathbf{F}^{1,\alpha^2}$.} 
    \label{fig13} 
\end{figure}

\begin{figure}[H]
    \centering
		{\includegraphics[width=0.3\linewidth]{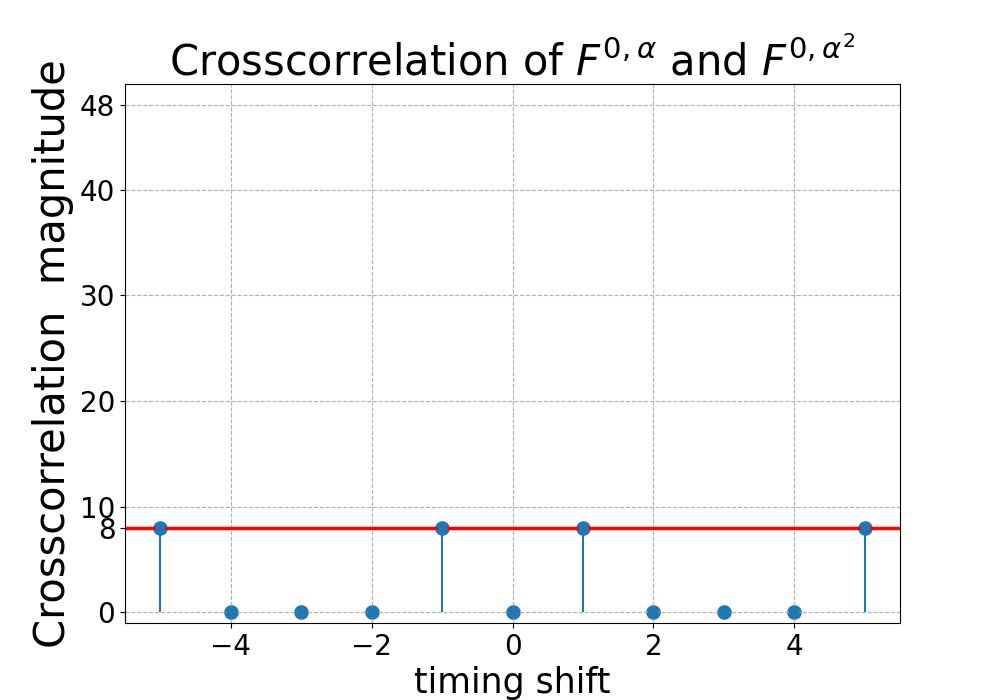}}
    \hfill
		{\includegraphics[width=0.3\linewidth]{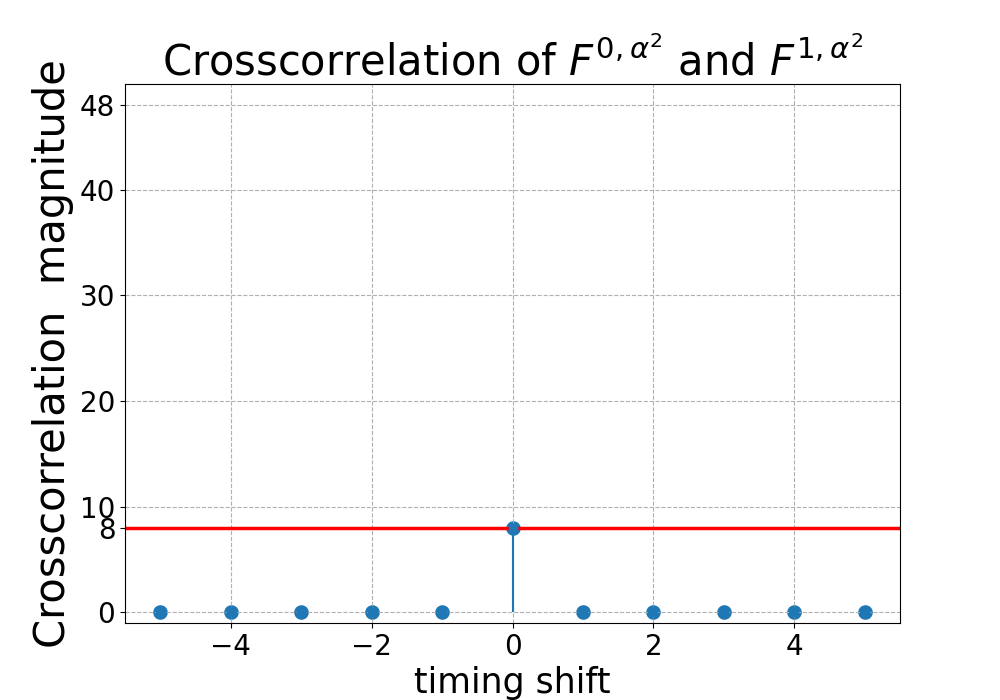}}
    \hfill
		{\includegraphics[width=0.3\linewidth]{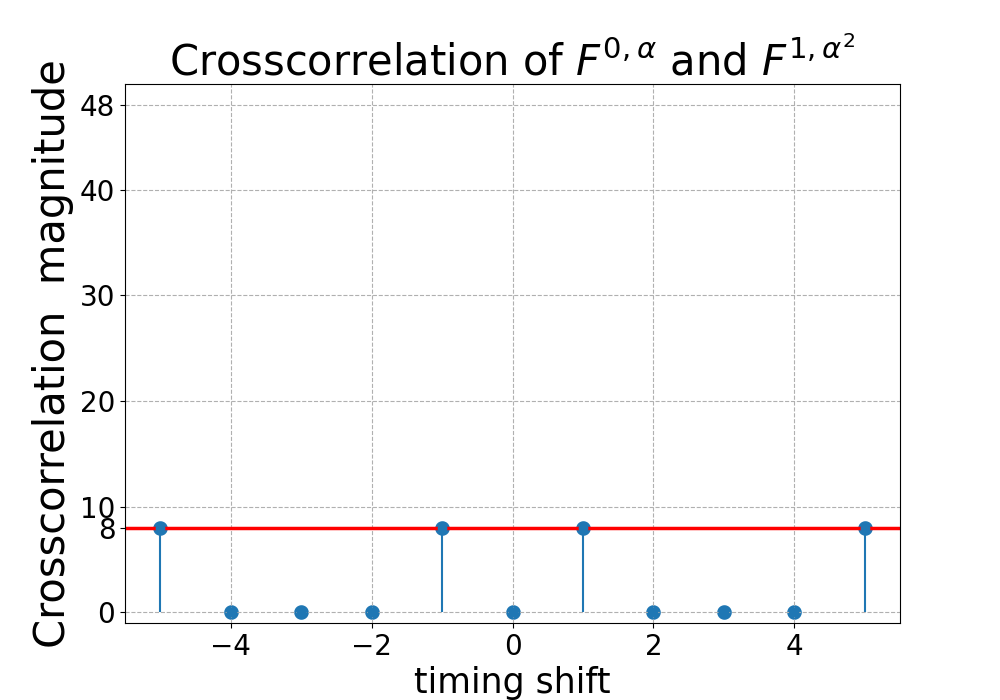}}
    \caption{The cross-correlation magnitude distributions of $\mathbf{F}^{0,\alpha}$,$\mathbf{F}^{0,\alpha^2}$ and $\mathbf{F}^{1,\alpha^2}$.} 
    \label{fig14} 
\end{figure}

\end{example}

\section{Concluding remarks}\label{sec6}
In this paper, we presented a new construction of periodic QCSSs and five new constructions of aperiodic QCSSs. The highlights of these QCSSs are as follows:
\begin{enumerate}
\item Each family of these QCSSs has a large set size $\Theta(K^2)$ for some flock size $K$;
\item Compared with known QCSSs, these  QCSSs have improved parameters or new  lengths of their constituent sequences, which was pointed out in Section I;
\item All of these QCSSs are asymptotically optimal with respect to the correlation lower bound in (\ref{zl1}) or (\ref{zl2}), and
all of the five families of aperiodic QCSSs have only two-valued correlation magnitude. Specially, the QCSS $\mathcal{F}$ has zero aperiodic auto-correlation magnitude for all  non-zero time shifts. 
\item Both of the QCSSs $\mathcal{C}$ and $\mathcal{B}$ have very small alphabet size $p$.
\end{enumerate}
Recently, Heng et al. constructed the first two families of near-optimal period QCSSs with set size $\Theta(K^3)$ for some flock size $K$ \cite{ZH1}.
It is open to construct asymptotically optimal or near-optimal aperiodic QCSSs with set size $\Theta(K^n)$ for some flock size $K$ and $n>2$.



\section*{Declarations}
\textbf{Conflict of interest} The authors declare that they have no conflicts of interest relevant to the content of this article.


\begin{thebibliography}{}
\providecommand{\url}[1]{#1}
\csname url@samestyle\endcsname
\providecommand{\newblock}{\relax}
\providecommand{\bibinfo}[2]{#2}
\providecommand{\BIBentrySTDinterwordspacing}{\spaceskip=0pt\relax}
\providecommand{\BIBentryALTinterwordstretchfactor}{4}
\providecommand{\BIBentryALTinterwordspacing}{\spaceskip=\fontdimen2\font plus
\BIBentryALTinterwordstretchfactor\fontdimen3\font minus
  \fontdimen4\font\relax}
\providecommand{\BIBforeignlanguage}[2]{{%
\expandafter\ifx\csname l@#1\endcsname\relax
\typeout{** WARNING: IEEEtran.bst: No hyphenation pattern has been}%
\typeout{** loaded for the language `#1'. Using the pattern for}%
\typeout{** the default language instead.}%
\else
\language=\csname l@#1\endcsname
\fi
#2}}
\providecommand{\BIBdecl}{\relax}
\BIBdecl

\end{thebibliography}


\begin{thebibliography}{1}
\bibliographystyle{IEEEtran}

\bibitem{ZZ2}
A. R. Adhikary, Y. Feng, Z. Zhou and P. Fan, ``Asymptotically optimal and near-optimal aperiodic quasi-complementary sequence sets based on florentine rectangles'',\textit{ IEEE Trans. Commun.}, vol. 70, no. 3, pp. 1475-1485, March 2022.

\bibitem{JA1}
J. A. Davis and J. Jedwab, ``Peak-to-mean power control
in OFDM Golay complementary sequences and Reed-Muller codes'', \textit{IEEE Trans. Inf. Theory}, vol.45, no. 7, pp.2397-2417, Nov 1999.

\bibitem{Cao1}
X. Cao, X. Gu, and J. Wan. ``A new kind of hybrid character sums and their applications''. in \emph{Proc. of 10th International Workshop on Signal Design and Its Applications in Communications (IWSDA)}, IEEE, 2022.


\bibitem{C1} H.-H. Chen, J.-F. Yeh, and N. Suehiro, ``A multicarrier CDMA architecture based on orthogonal complementary codes for new generations
of wideband wireless communications,'' \emph{IEEE Commun. Mag.}, vol. 39,
no. 10, pp. 126-135, 2001.


\bibitem{F} P. Fan, W. Yuan, and Y. Tu, ``Z-complementary binary sequences,'' \textit{IEEE
 Signal Process. Lett.}, vol. 14, no. 8, pp. 509-512, Aug. 2007.


\bibitem{G}
M. Golay, ``Complementary series,'' \textit{IRE Trans. Inf. Theory}, vol.7, no.2, pp. 82-87, Apr. 1961.


\bibitem{ZH1}
Z. Heng, P. Wang, C. Xie, H. Zhou, ``Large sets of quasi-complementary sequences from polynomials over finite fields and Gaussian sums,'' \textit{arXiv}:2411.04445v2, 2024. 

\bibitem{LY1}
Y. Li, T. Liu, and C. Xu, ``Constructions of asymptotically optimal quasi-complementary sequence sets,'' \textit{IEEE Commun. Lett.}, vol. 2, no. 8, pp. 1516-1519, Aug 2018.

\bibitem{LY2}
Y. Li, L. Tian, T. Liu, and C. Xu, ``Two constructions of asymptotically
optimal quasi-complementary sequence sets,'' \textit{IEEE Trans. Commun.},
vol. 67, no. 3, pp. 1910-1924, 2019.

\bibitem{LY3}
Y. Li, L. Tian, T. Liu, and C. Xu, ``Constructions of quasi-complementary
sequence sets associated with characters,'' \textit{IEEE Trans. Inf. Theory}, vol. 65, no. 7, pp. 4597-4608, 2019.

\bibitem{LY4}
Y. Li, T. Yan, and C. Lv, ``Construction of a near-optimal quasi-complementary sequence set from almost difference set,'' \textit{Cryptogr. Commun.}, vol. 11, no. 4, pp. 815-824, 2019.

\bibitem{LY5}
Y. Li, L. Tian and C. Xu, ``Constructions of asymptotically optimal aperiodic quasi-complementary sequence sets,'' \textit{IEEE Trans. Commun.}, vol. 67, no. 11, pp. 7499-
7511, Nov 2019.
 
\bibitem{LG}
G. Luo, X. Cao, M. Shi, and T. Helleseth, ``Three new constructions of asymptotically optimal periodic quasi-complementary sequence sets with small alphabet sizes,'' \textit{IEEE Trans. Inf. Theory}, vol. 67, no. 8, pp. 5168-5177, 2021.

\bibitem{Lidl}
R. Lidl, H. Niederreiter, \emph{Finite Fields}, Cambridge Univ. Press, Cambridge, U.K., 1997.


\bibitem{Liu} Z. Liu, ``Perfect-and quasi-complementary sequences,'' Ph.D. dissertation, School Elect. Electron. Eng., Nanyang Technol. Univ., Singapore, 2014.

\bibitem{ZL}
Z. Liu, U. Parampalli, Y. L. Guan, and S. Boztas, ``Constructions of optimal and near-optimal quasi-complementary sequence sets from singer difference sets,'' \textit{IEEE Wireless Commun. Lett.}, vol. 2, no. 5, pp. 487-490, 2013.

\bibitem{ZL1}
Z. Liu, Y. L. Guan, and W. H. Mow, ``A tighter correlation lower bound for quasi-complementary sequence sets,''\textit{IEEE Trans. Inf. Theory}, vol. 60, no. 1, pp. 388-396, Jan 2014.

\bibitem{ZL2}
Z. Liu, Y. L. Guan, B. C. Ng, and H. H. Chen, ``Correlation and set size bounds of complementary sequences with low correlation zones,'' \textit{IEEE Trans. Commun.}, vol. 59, no. 12, pp. 3285-3289, Dec 2011.

\bibitem{ZL3}
Z. Liu, Y. L. Guan, and, W. H. Mow, ``Asymptotically locally optimal weight vector design for a tighter correlation lower bound of quasi-complementary sequence
sets,'' \textit{IEEE Trans. Signal Process.}, vol. 65, no. 12, pp.  3107-3119, Jun 2017.

\bibitem{ZL4}
Z. Liu, Y. L. Guan and W. H. Mow, ``Improved lower bound for quasi-complementary sequence set,'' \textit{Proc. IEEE Int. Symp. Inf. Theory}, St. Petersburg, pp. 489-493, 2011.

\bibitem{AP1}
A. Pezeshki, A. R. Calderbank, W. Moran and S. D. Howard, ``Doppler resilient Golay complementary waveforms,'' \textit{IEEE Trans. Inf. Theory}, vol. 54, no. 9, pp. 4254-4266, Sep. 2008.
 


\bibitem{PS1}
P. Spasojevic and C. N. Georghiades, ``Complementary sequences for ISI channel estimation,'' \textit{IEEE Trans. Inf. Theory}, vol. 47, no. 3, pp. 1145-1152, 2001.

\bibitem{MKS}
M. K. Simon, J. K. Omura, R. A. Scholtz, and B. K. Levitt, ``Spread
Spectrum Communications,'' Rockville, MD, USA: Computer Science, 1985.

\bibitem{ZZ3}
B. Shen, T. Yu, Z. Zhou  and Y. Yang. ``Asymptotically optimal aperiodic quasi-complementary sequence sets based on extended Boolean functions,'' \textit{Des. Codes Cryptogr}, vol. 92, pp. 4213-4230, 2024.

\bibitem{PS2}
P. Sarkar, C. Li, S. Majhi and Z. Liu, ``New correlation bound and construction of quasi-complementary sequence sets,'' \textit{IEEE Trans. Inf. Theory}, vol. 70, no. 3, pp. 2201-2223, March 2024.

\bibitem{T}
X. Tang, P. Fan, D. Li, N Suehiro, ``Binary array set with zero correlation zone,''\textit{ Electronic Letters}, vol.37, No.13, pp.841-842, 2001.

\bibitem{SW1}
S. Wang and A. Abdi, ``MIMO ISI channel estimation using uncorrelated Golay complementary sets of polyphase sequences,'' \textit{IEEE Trans. Veh. Technol.}, vol. 56, no. 5, pp. 3024-3039, Sep 2007.

\bibitem{WM1}
M. Wang, T. Liu, M. Liu, Y. Li, ``Two Constructions of Aperiodic Quasi-Complementary Sequence Sets With New Parameters,'' \textit{IEEE Trans on Commun.}, 2024.

\bibitem{Welch}
L. Welch, ``Lower bounds on the maximum cross correlation of signals (Corresp.),'' \textit{IEEE Trans. Inf. Theory}, vol. 20, no. 3, pp. 397-399, 1974.

\bibitem{XLC}
H. Xiao, G. Luo, X. Cao, ``New constructions of asymptotically optimal quasi-complementary sequence sets with small alphabet sizes.'' \textit{IEEE Trans. on Commun.}, DOI: 10.1109/TCOMM.2024.3524020, 2024.

\bibitem{ZZ1}
Z. Zhou, F. Liu, A. R. Adhikary and P. Fan, ``A generalized construction of multiple complete complementary codes and asymptotically optimal aperiodic quasi-complementary sequence sets,''\textit{IEEE Trans. Commun.}, vol. 58, no. 6, pp. 3564-3571, 2020.

\end{thebibliography}
\end{document}